\documentclass[epsf,useAMS,usenatbib, usegraphicx]{mn2e}
\usepackage{graphicx}
\usepackage{epsfig}
\usepackage{amsmath}
\usepackage{amssymb}
\usepackage{rotating}
\usepackage{color}
\usepackage{pifont}
\usepackage{longtable}
\usepackage{pdflscape,lscape}
\usepackage{float}
\usepackage{placeins}
\newcommand{\ion}[2]{{#1}\,{\sc #2}}
\newcommand{\teff}{$T_{\rm eff}$}
\newcommand{\logg}{$\log g$}
\newcommand{\vsini}{$v \sin i$}
\newcommand{\kms}{km\,s$^{-1}$}
\usepackage{appendix}
\usepackage{booktabs}
\usepackage{upgreek}
\usepackage{url}

\usepackage{natbib}
\title[Detailed studies of detached eclipsing binaries]{Detailed spectroscopic and photometric study of three detached eclipsing binaries}
\author[F. Kahraman Ali\c{c}avu\c{s} \& F. Ali\c{c}avu\c{s}]{F. Kahraman Ali\c{c}avu\c{s}$^{1,2}$\thanks{E-mail: filizkahraman01@gmail.com},
F. Ali\c{c}avu\c{s}$^{2}$
\\
$^{1}$Nicolaus Copernicus Astronomical Center, Bartycka 18, PL-00-716 Warsaw, Poland\\ 
$^{2}$Canakkale Onsekiz Mart University, Faculty of Sciences and Arts, Physics Department, 17100, Canakkale, Turkey\\
}
\begin{document}

\date{Accepted ... Received ...; in original form ...}

\pagerange{\pageref{firstpage}--\pageref{lastpage}} \pubyear{2019}

\maketitle

\label{firstpage}

\begin{abstract}

Detached eclipsing binaries are remarkable systems to provide accurate fundamental stellar parameters. The fundamental stellar parameters and the metallicity values of
stellar systems are needed to deeply understand the stellar evolution and formation. In this study, we focus on the detailed spectroscopic and photometric studies of three detached
eclipsing binary systems, V372\,And, V2080\,Cyg, and CF\,Lyn to obtain their accurate stellar, atmospheric parameters,
and chemical compositions. An analysis of light and radial velocity curves was
carried out to derive the orbital and stellar parameters. The disentangled spectra of component stars were obtained for the spectroscopic analysis.
Final \teff, \logg, $\xi$, \vsini\, parameters and the element abundances of component stars were derived by using the spectrum synthesis method. The
fundamental stellar parameters were determined with a high certainty for V372\,And, V2080\,Cyg ($\sim$$1-2$\%) and with an accuracy for CF\,Lyn ($\sim$$2-6$\%).
The evolutionary status of the systems was examined and their ages were obtained. It was found that the component stars 
of V2080\,Cyg have similar iron abundance which is slightly lower than solar iron abundance. Additionally, we showed that
the primary component of CF\,Lyn exhibits a non-spherical shape with its 80\% Roche lobe filling factor. 
It could be estimated that CF\,Lyn will start its first Roche overflow in the next 0.02\,Gyr.

\end{abstract}


\section{Introduction}

Eclipsing binary stars are one of the most important sources for astrophysics to understand the evolution and the formation 
of stars. They are valuable variables which may supply the fundamental stellar parameters (e.g. mass and radius) with an accuracy 
of about 1\% \citep{2015ApJ...802..102L, 2013A&A...557A.119S}. 
The orbital parameters of an eclipsing binary system can be obtained by the 
analysis of radial velocity curve. The orbital inclination and the radii of component stars according to the 
orbital semi-major axis can only be determined by the light curve analysis of eclipsing binary systems. Therefore,
a simultaneous analysis of the high-quality radial velocity and light curves is essential for deriving the fundamental stellar 
parameters from eclipsing binary stars. Precise 
fundamental stellar parameters offer us a possibility to investigate the stellar evolution in detail and to test the present 
evolutionary models. 
% Therefore the eclipsing binary stars are remarkable objects for astrophysicists.

% The eclipsing binary stars are generally classified to be detached, semi-detached, and contact eclipsing binary systems by 
% taking into account their Roche geometries. Each of these eclipsing binary types exhibits different physical phenomena. Although all these 
% types of eclipsing binary stars supply the fundamental stellar parameters, the detached double-lined eclipsing binary variables offer 
% a better accuracy in the fundamental stellar parameters comparing to the other eclipsing binary types \citep[][e.g.]{2010A&ARv..18...67T}.

In a theoretical evolutionary model of a star, mass is a basic parameter. 
In addition to this parameter, metallicity has also a significant effect on the evolution. Therefore, 
to calculate an accurate theoretical evolutionary model, it is necessary to obtain a precise value of both mass and metallicity. 
In the case of double-lined eclipsing binary stars, masses of component stars can be derived from the analysis of light and radial velocity curves, 
whereas the accurate metallicity parameter can be determined from the spectroscopic analysis of individual spectra of component stars. 
For this purpose, the spectral disentangling method was developed \citep{1994A&A...281..286S}. This method has been applied by several 
authors \citep[][e.g.]{2018MNRAS.481.3129P, 2018MNRAS.481.5660D, 2015AJ....150...55S} 
to obtain the individual spectra of component stars in double-lined eclipsing binaries. By performing this method, 
the metallicity value, as well as the atmospheric parameters of each component star can be derived separately. 

The most recent eclipsing binary catalogue was published by \citet{2013AN....334..860A}. The catalogue consists of 7200 stars including detached, 
semi-detached and contact binaries.
A catalogue for detached double-lined eclipsing binary stars was given by \citet{2014PASA...31...24E}. In this catalogue, 
they presented the range of effective temperature (\teff), mass ($M$), and radius ($R$) of those 
eclipsing binaries and showed that for these binaries the accuracy of $M$ and $R$ measurements changes from 1\% to 5\%. 
Additionally, they pointed out the missing sample of giant and supergiant component stars inside their catalogue. The other 
catalogue of detached eclipsing binaries is ``DEBcat: catalogue of the physical properties of well-studied eclipsing 
binaries''\footnote{http://www.astro.keele.ac.uk/jkt/debcat/} \citep{2015ASPC..496..164S}. 
This catalogue has been available since 2006 and it 
is a continuation of the list given by \citet{1991A&ARv...3...91A}. The catalogue contains precise 
fundamental stellar parameters ($M$, $R$), 
surface gravity (\logg), \teff , luminosity ($L$), and metallicity values of detached systems. The DEBcat is always updated when a new 
precise measurement of detached eclipsing binaries is presented in the literature.

To improve the stellar evolutionary models and examine the star formation in detail, the double-lined eclipsing binaries, in particular, the detached ones, 
are very valuable. Therefore, in this study, we focus on an extensive spectroscopic and photometric analysis of three detached eclipsing binary stars, 
V2080\,Cyg, V372\,And, and CF Lyn. As these systems have enough archival data, they were were selected for this study.
All systems were first classified to be variable stars by the Hipparcos \citep{1997ESASP1200.....E}. V2080\,Cyg (HIP\,95611) has a spectral type of F4V\,+\,F4V 
\citep{2014PASA...31...24E}. The first detail study of V2080\,Cyg was published by \citet{2008MNRAS.384..331I}. 
In this study, they presented a radial velocity and 
light curve analysis of the binary system and classified it as a double-lined detached eclipsing binary. They also examined the pulsation variability 
in the component stars, however, no pulsation feature was found. V372\,And (HIP\,9740) is classified as a main-sequence detached system 
in the catalogue of eclipsing binary variables \citep{2013AN....334..860A}. The star has no detailed analysis in the literature. CF\,Lyn (HIP\,37748) is
also classified as a detached eclipsing binary system with a spectral type of F8 in the same eclipsing binary stars catalogue \citep{2013AN....334..860A}. 
The first light curve analysis of CF\,Lyn was presented by \citet{2017NewA...53...53Z}. 
No spectroscopic analysis is available for this star. 

To derive the fundamental stellar ($M$, $R$), atmospheric parameters (\teff, \logg\, and microturbulence velocity $\xi$) and the metallicity of all component stars, we carried out an extensive 
analysis of the selected eclipsing binary systems. We give the information about the used photometric and spectroscopic data in Sect.\,2. The detailed spectroscopic 
analysis including the radial velocity, spectral disentangling and the determination of atmospheric parameters is introduced in Sect.\,3. 
The light curve analysis of the systems is presented in Sect.\,4. The investigation of evolutionary status of the detached eclipsing binaries 
and an argument about the derived parameters are given in Sect.\,5. The conclusions are presented in Sect.\,6. 

\section[]{Photometric and spectroscopic data}

Photometric and spectroscopic data of the systems were collected from the public databases of the Super Wide Angle Search for Planets
(SuperWASP) and ELODIE, respectively.

The SuperWASP is a two-site ground-based programme which aims to discover exoplanets by transit method \citep{2006PASP..118.1407P}.
The SuperWASP observations have been carried out by the broad-band filters ($\sim$ 4000\,$-$\,7000\,\AA) since 2006. In addition to
exoplanet discoveries, the SuperWASP has provided data of many variable stars. The selected detached eclipsing binary stars also have
the data in the SuperWASP archive\footnote{https://wasp.cerit-sc.cz/form}. The data of the systems are sufficient for planning
photometric analysis. All available data in the database were collected. The scattered points beyond the 3$\sigma$ level were removed for the following analysis.
The available data points of the systems in the SuperWASP archive are given in Table\,1.

The spectra of the systems were taken from the ELODIE archive\footnote{http://atlas.obs-hp.fr/elodie/}. The ELODIE is an \'{e}chelle spectrograph
mounted on the 1.93-m telescope in the observatoire de Haute (France). The spectrograph was used between $\sim$ 1993 and 2006. It provided
spectra with a resolving power of 42000 and in a wavelength range of $\sim$ 3850 to 6800\,\AA. The ELODIE spectra are reduced automatically
by the dedicated pipeline. In this study, we used the reduced ELODIE spectra of the systems and normalised them by using the \textit{continuum} task of the
NOAO/IRAF\footnote{http://iraf.noao.edu/} package. The number of the available spectra of each binary system and the minimum and
maximum signal-to-noise (S/N) ratios of the spectra are given in Table\,1.

\section{Spectroscopic analysis}

A detailed spectroscopic analysis of the selected systems was planned to carry out to obtain the orbital parameters, 
the fundamental atmospheric parameters, abundances and the projected rotational velocity (\vsini) values of component 
stars in each detached eclipsing 
binary system. Therefore, the analysis of the radial velocity, the spectral disentangling, the atmospheric parameters, and the abundance was performed.

\begin{table}
\centering
\caption{Information for the selected eclipsing binaries.}
  \label{table1}
\begin{tabular*}{0.92\linewidth}{@{\extracolsep{\fill}}lcccc}
\hline
  Name      & V     & SuperWASP      &  ELODIE           &  S/N           \\
            & (mag) & Points         & Spectra           &  Range \\
\hline 
V2080 Cyg   & 7.90  & 5858          &  8                 & 20\,-\,140 \\
V372 And    & 9.28  & 2718          &  10                & 25\,-\,80 \\
CF Lyn      &10.05  & 5461          &  10                & 20\,-\,75 \\
\hline
% \bottomrule
\end{tabular*}
\label{abunresult}
\end{table}

\subsection{Radial velocity measurements and analysis}
To derive the accurate orbital parameters of eclipsing binary systems via the radial velocity analysis, the measured radial
velocities of component stars are preferred to be spread over the orbital phase. In our study, the used spectra offer this property.
To obtain the radial velocity ($v$$_{r}$) measurements of each component star in the binaries, we used the FXCOR task of the NOAO/IRAF package
program. In the measurements, the radial velocity standard star HD\,50692 ($v$$_{r}$\,=\,-15.5\,\kms, 
\citeauthor{1999ASPC..185..367U} \citeyear{1999ASPC..185..367U})
was used as a template. During the analysis, the spectra having the lowest S/N ratio in each system were excluded.
The obtained $v$$_{r}$ measurements are given in Table\,\ref{rvmeasuremnt}.

\begin{figure*}
 \centering
 \label{rvfitss}
 \begin{minipage}[b]{0.33\textwidth}
  \includegraphics[height=6cm, width=1\textwidth]{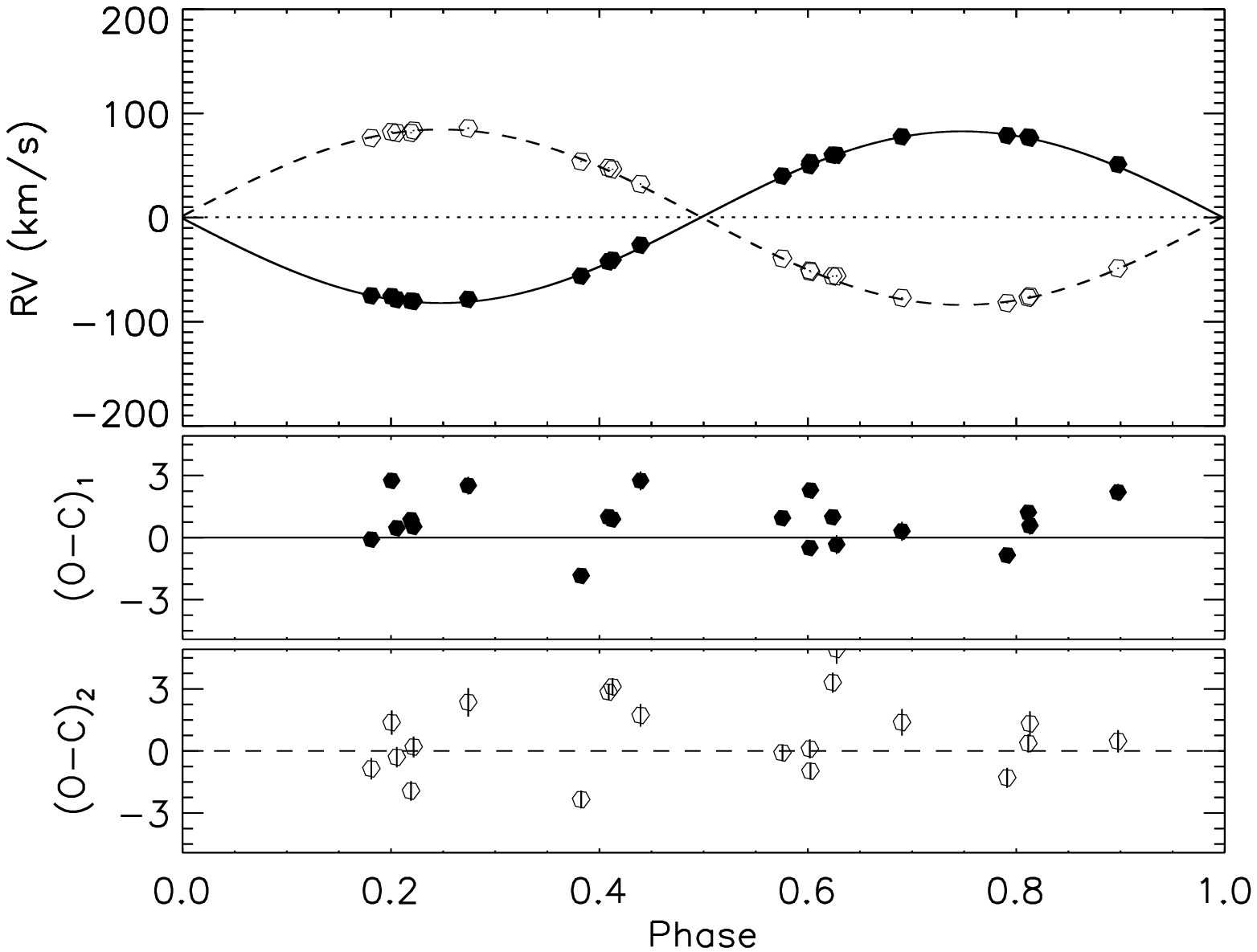}
%   \caption{1}
 \end{minipage}
 \begin{minipage}[b]{0.33\textwidth}
  \includegraphics[height=6cm, width=1\textwidth]{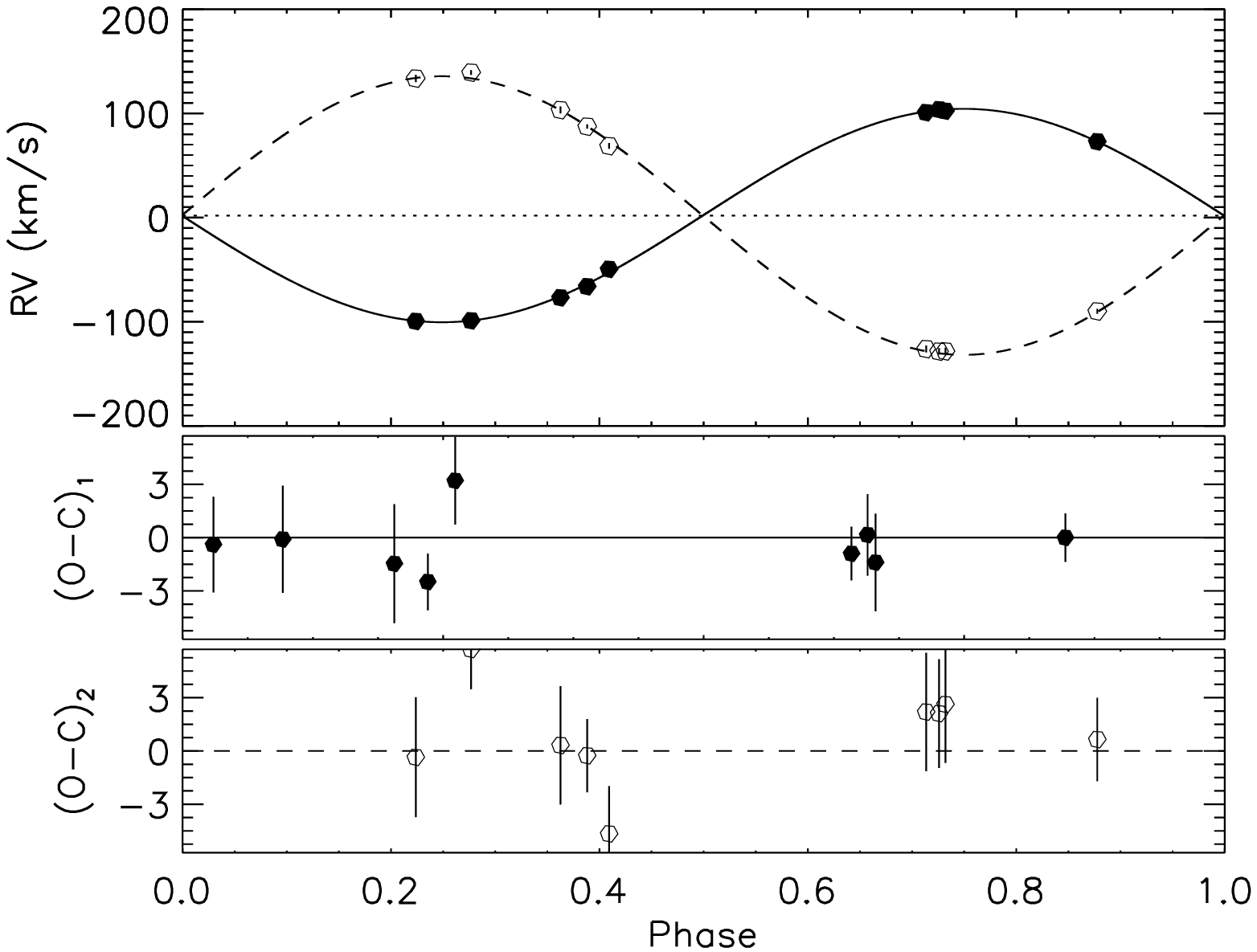}
  \end{minipage}
 \begin{minipage}[b]{0.33\textwidth}
  \includegraphics[height=6cm, width=1\textwidth]{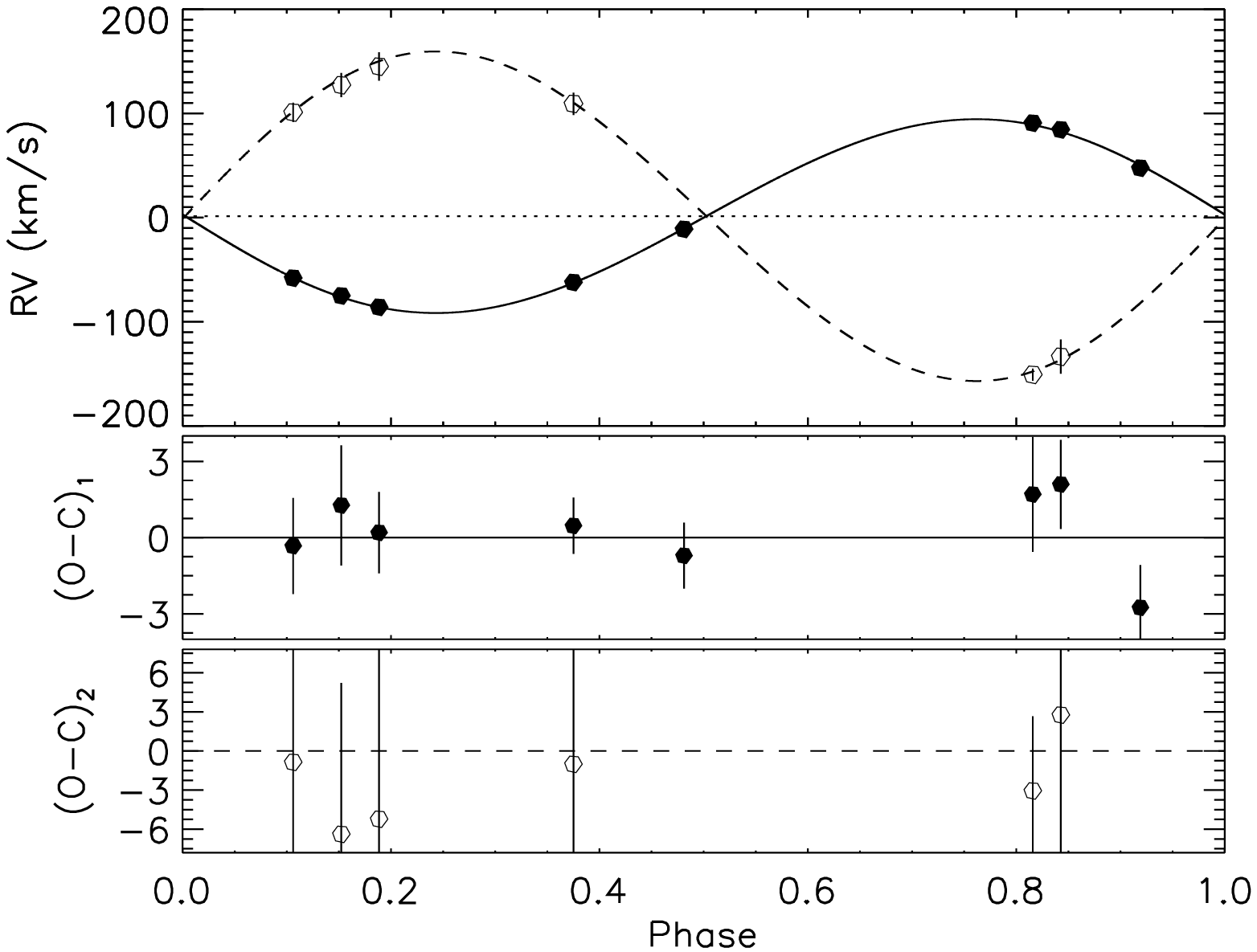}
%   \caption{3}
 \end{minipage}
   \caption{The theoretical $v$$_{r}$ curves fit for the measured $v$$_{r}$ values (upper panels) and the residuals (O-C) in \kms (lower panels).
   The left, middle and right panels belong to V2080 Cyg, V372 And, and CF Lyn, respectively. The subscripts 1 and 2 represent the primary and the secondary components, respectively.}
\end{figure*}

We derived the orbital parameters, eccentricity $e$, argument of periastron $\omega$, the velocity of the system mass center
V$\gamma$, mass ratio $q$ ($M_\text{secondary}/M_\text{primary}$), $a \sin i$ ($a$ is semi-major axis and $i$ is inclination),
and the amplitude of the $v$$_{r}$ curve of the star with respect to the center of mass of the binary $K$ by analysing
the determined $v$$_{r}$ curves of the systems. In the analysis, we used the \texttt{rvfit} code\footnote{http://www.cefca.es/people/riglesias/rvfit.html}.
The \texttt{rvfit} code fits the $v$$_{r}$ curves of double-lined, single-lined binaries or exoplanets by using a minimization
method of adaptive simulated annealing \citep{2015PASP..127..567I}. It is a user-friendly code which automatically fits the
given orbital parameters above, the orbital period $P$ and also the periastron passage time $T_{p}$.

In the $v$$_{r}$ analysis, the $P$ and $T_{p}$ parameters of the binaries were kept fixed and they were taken from \citet{2004AcA....54..207K}.
In the analysis of V2080\,Cyg, we also used the previously obtained $v$$_{r}$ measurements by
\citet{2008MNRAS.384..331I}. As a result of the analysis, the orbital parameters were derived and
it was found that V2080\,Cyg and V372\,And have a circular orbit, whereas CF\,Lyn
has an eccentric orbit with a value of about 0.03. The uncertainties of the obtained result were calculated using the
Markov Chain Monte Carlo method. The results are listed in Table\,\ref{rvresult}. The theoretical $v$$_{r}$ curves fit to the
obtained $v$$_{r}$ measurements and the difference between them are shown in Fig.\,1.

\begin{table*}
\begin{center}
\centering
\caption[]{The results of the radial velocity analysis. The subscripts 1 and 2 represent the primary and the secondary components, respectively.
$^a$ shows the fixed parameters.}\label{rvresult}
\begin{tabular*}{0.7\linewidth}{@{\extracolsep{\fill}}lrrr}
% \begin{tabular}{lccc}
\hline
 Parameter		 &  V2080 Cyg	        &V372 And                       & CF Lyn              \\
\hline
$T_p$ (HJD)$^{a}$	 & 2452504.186                  &2452501.191                    &2452500.070      \\
$P$ (d)$^{a}$            & 4.9335660	                &2.9410200                      &1.3853720           \\
$\gamma$ (km/s)	         & 1.07\,$\pm$\,0.07	&1.96 $\pm$ 0.54       &1.36 $\pm$ 0.73      \\
$K_1$ (km/s)		 & 82.35 $\pm$ 0.10	&102.43 $\pm$ 0.95     &93.03 $\pm$ 1.15   \\
$K_2$ (km/s)		 & 83.99 $\pm$ 0.16	&133.72 $\pm$ 1.21     &158.22 $\pm$ 4.68    \\
$e$			 & 0.000 $\pm$ 0.001	&0.000 $\pm$ 0.004     &0.030$\pm$ 0.009  \\
$\omega$ (deg)		 & 90.63 $\pm$ 0.07     &90.23 $\pm$ 0.62      &89.09 $\pm$ 0.58  \\
$a_1\sin i$ ($10^6$ km)	 & 5.587 $\pm$ 0.007    &4.142 $\pm$ 0.038     &1.771 $\pm$ 0.022  \\
$a_2\sin i$ ($10^6$ km)	 & 5.698 $\pm$ 0.011    &5.408 $\pm$ 0.049     &3.013 $\pm$ 0.089   \\
$a  \sin i$ ($10^6$ km)	 & 11.285 $\pm$ 0.013   &9.550 $\pm$ 0.062     &4.784 $\pm$ 0.091  \\
$M_1\sin ^3i$ ($M_\odot$)& 1.188 $\pm$ 0.005	&2.272 $\pm$ 0.048     &1.432 $\pm$ 0.097      \\
$M_2\sin ^3i$ ($M_\odot$)& 1.165 $\pm$ 0.003	&1.740 $\pm$ 0.035     &0.842 $\pm$ 0.036     \\
$q = M_2/M_1$		 & 0.980 $\pm$ 0.002    &0.766 $\pm$ 0.010     &0.588 $\pm$ 0.019   \\
 \hline
\end{tabular*}
     \end{center}
\end{table*}

\subsection{Spectral disentangling}

For the investigation of the evolutionary status of each component star in the selected detached eclipsing binaries, the metallicity 
values are needed to be known. However, the spectra of the systems are composite. Therefore, to obtain the spectra of individual 
component stars in the binaries, we applied the spectral disentangling method. The \texttt{FDBINARY} code \citep{2004ASPC..318..111I} was used in 
the analysis. The \texttt{FDBINARY} disentangles a composite spectrum in the Fourier space. It is also capable of resolving the 
third component spectrum. To apply the disentangling method via the \texttt{FDBINARY} first light contributions of component stars 
in total should be known. Therefore, we first carried out a preliminary light curve analysis by assuming the primary 
component stars' \teff\, values from their spectral type. In this assumption, the spectral type\,$-$\,\teff\, relation given by 
\citet{2009ssc..book.....G} was used. As a result of this preliminary analysis, we obtained the light contributions of each
component star in total. It turned out that V372\,And and CF\,Lyn are two-body systems, while V2080\,Cyg has a third component.
Approximately, the light contributions of the primary and secondary components were found to be
77\% and 23\% for V372\,And, 94\% and 6\% for CF\,Lyn. The light contributions of the primary, secondary and third components of V2080\,Cyg
were also obtained about 52\%, 40\%, and 8\%, respectively. The updated light curve
analysis will be given in Sect.\,4.

In the application of the spectral disentangling method, we used the spectral interval of ~4000$-$5500\,\AA. This spectral 
interval was divided into ten parts with 100$-$150\,\AA\, steps, and each 
part was analysed separately. In the analysis, the spectra having the lowest S/N ratios were not used. 
The derived orbital parameters were used as inputs. The values of $P$ and $\omega$ were fixed, 
while $T_p$, $K_{1}$, $K_{2}$, and $e$ were kept as free parameters in the analysis. 
After the disentangled spectra of each spectral part were obtained, 
they were re-normalised considering the light ratio of components obtained from the initial light curve analysis. In this process, the procedure given by 
\citet{2004ASPC..318..111I} was used.

As a result of this analysis, a disentangled spectrum with a higher S/N ratio than the average S/N 
ratio of the input spectra can be obtained \citep{2009MNRAS.394.1519P}. The final S/N ratios 
of the disentangled spectra of each star were calculated by using the given equation by \citet{2009MNRAS.394.1519P}. 
The final S/N ratios for the primary, secondary and the third components of V2080\,Cyg were found to be 135, 103 and 21, respectively. 
For V372\,And, those values were calculated as 157 for primary and 50 for secondary components. The final S/N values were also 
obtained to be 140 and 9 for the primary and secondary components of CF\,Lyn. As can be seen from the result of S/N ratios, 
except for the third component of V2080\,Cyg and the secondary component of CF\,Lyn, all disentangled spectra are sufficient for 
the atmospheric parameter determination. Additionally, they are good enough for the abundance analysis. Only the secondary 
component of V372\,And could give results with higher errors because of its lower S/N ratio.

\begin{figure*}
 \centering
 \label{hlines}
 \begin{minipage}[b]{0.33\textwidth}
  \includegraphics[height=6cm, width=1\textwidth]{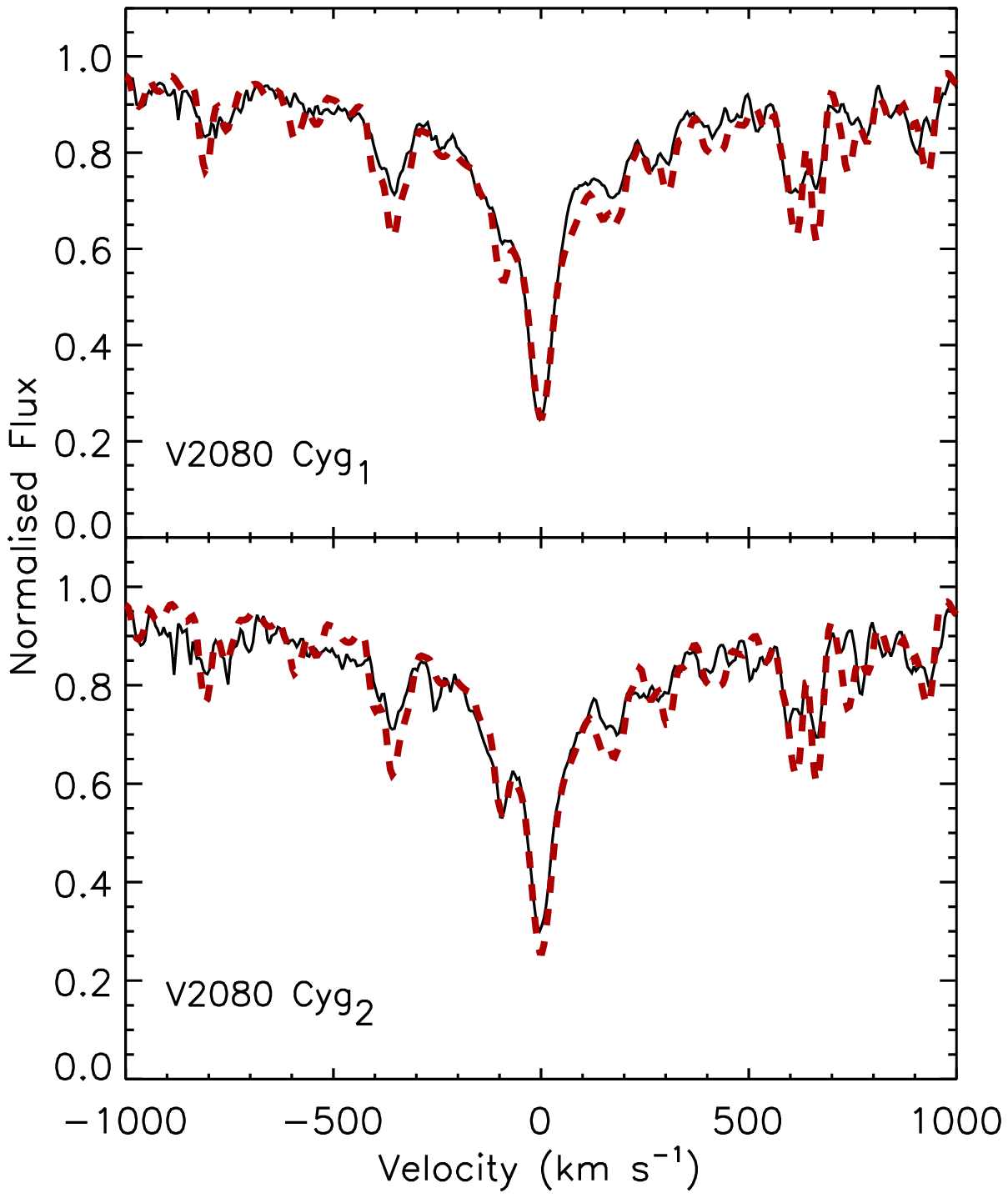}
%   \caption{1}
 \end{minipage}
 \begin{minipage}[b]{0.33\textwidth}
  \includegraphics[height=6cm, width=1\textwidth]{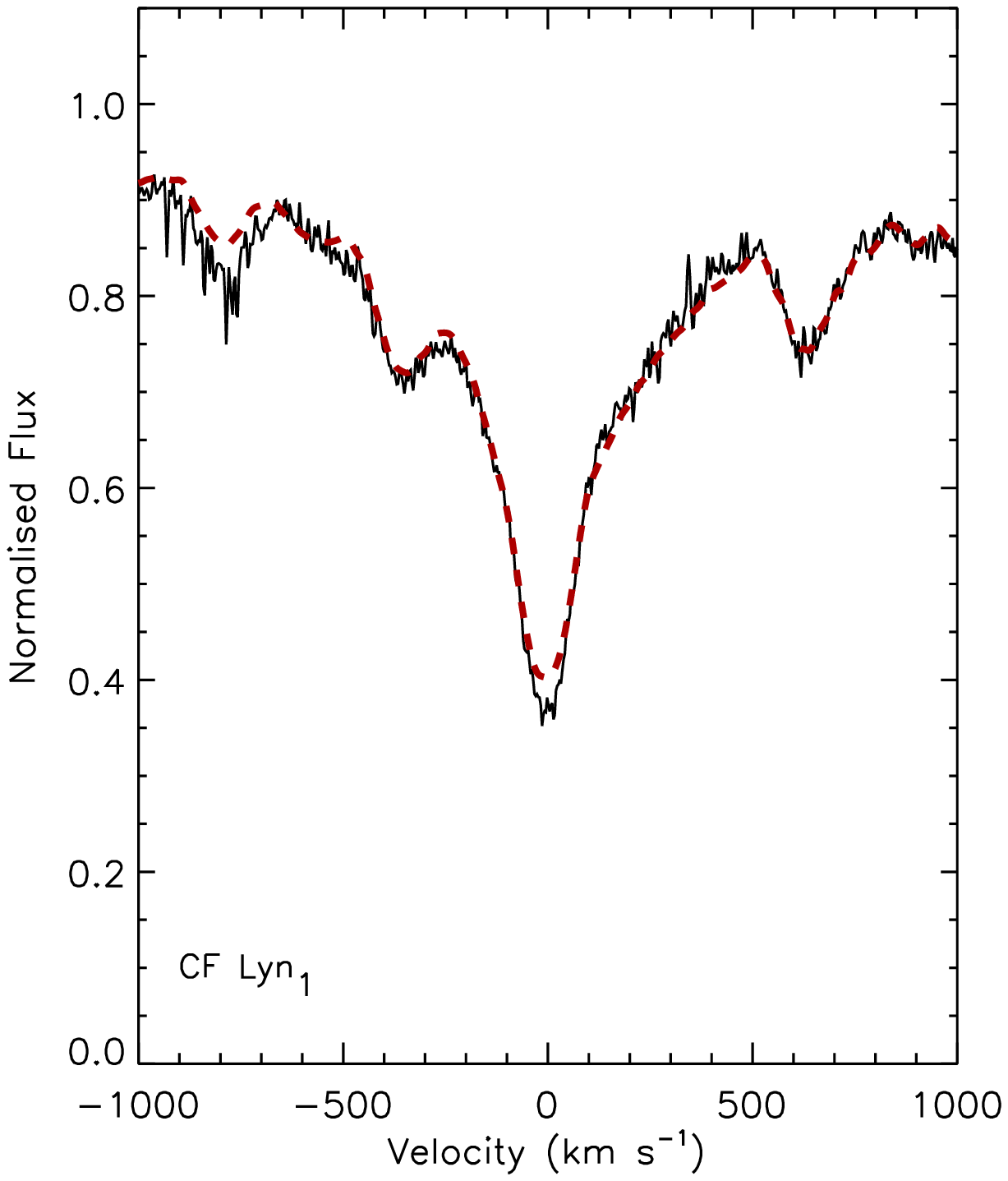}
  \end{minipage}
 \begin{minipage}[b]{0.33\textwidth}
  \includegraphics[height=6cm, width=1\textwidth]{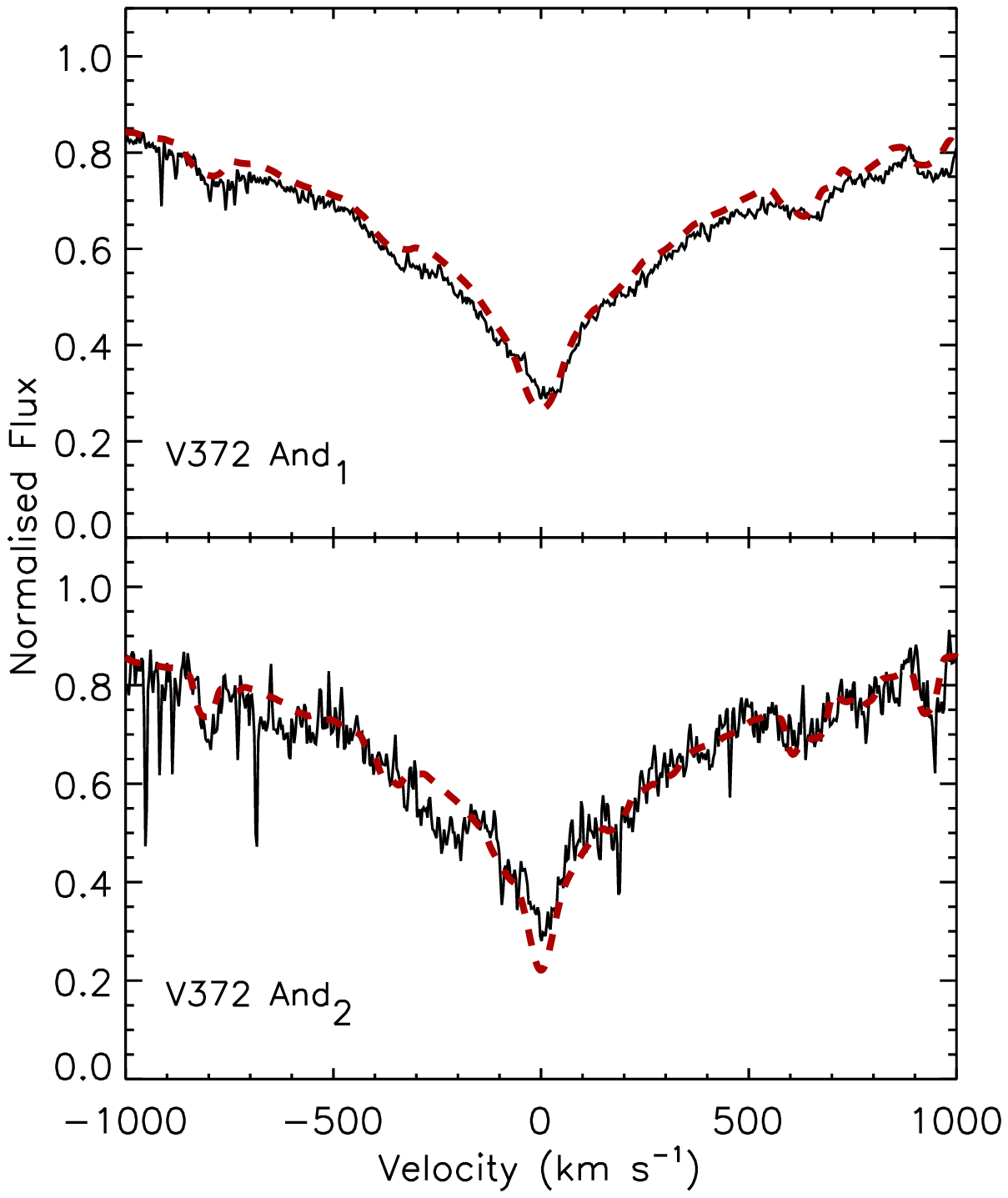}
%   \caption{3}
 \end{minipage}
   \caption{Comparison of the theoretical (dashed line) and disentangled $H_{\beta}$ lines (continuous line) of component stars.
  The subscripts 1 and 2 represent the primary and the secondary components, respectively.}
\end{figure*}

\subsection{Determination of the atmospheric parameters and the chemical abundances of component stars}

The disentangled spectra of each component star were analysed to determine the atmospheric parameters and the 
abundances of the stars. Before the determination of element abundances, accurate atmospheric parameters should be obtained. 
Therefore, we first derived the fundamental atmospheric parameters \teff, \logg, $\xi$, and also \vsini\, values of each component star. 

In this analysis, we used the plane-parallel, hydrostatic, line-blanketed local thermodynamic equilibrium (LTE) ATLAS9 model atmospheres 
\citep{1993KurCD..13.....K}. 
The theoretical spectra were synthesized with the SYNTHE code \citep{1981SAOSR.391.....K}.

\begin{figure*}
 \centering
 \label{spectrafit1}
 \begin{minipage}[b]{0.48\textwidth}
  \includegraphics[height=6cm, width=1\textwidth]{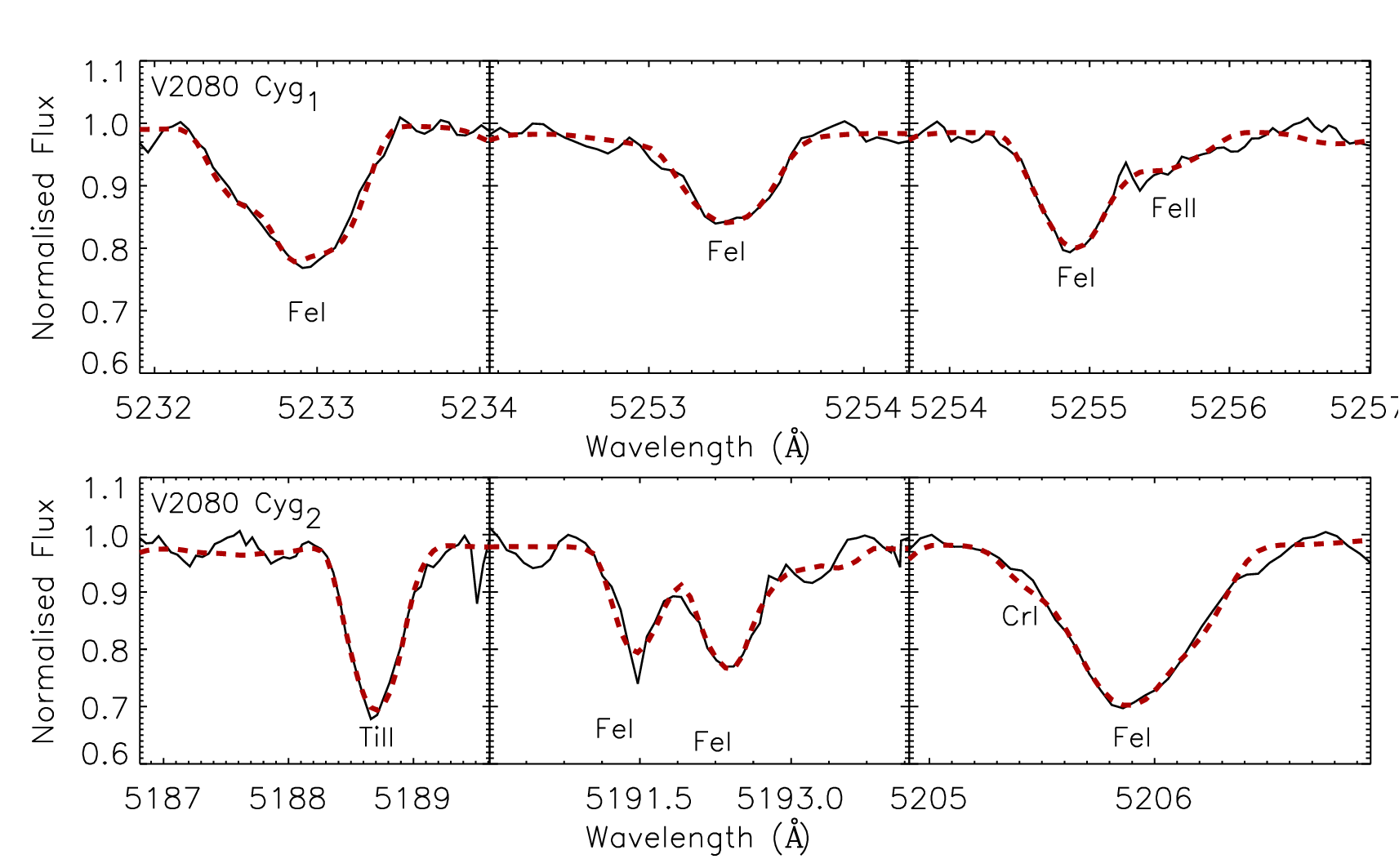}
%   \caption{1}
 \end{minipage}
 \begin{minipage}[b]{0.48\textwidth}
  \includegraphics[height=6cm, width=1\textwidth]{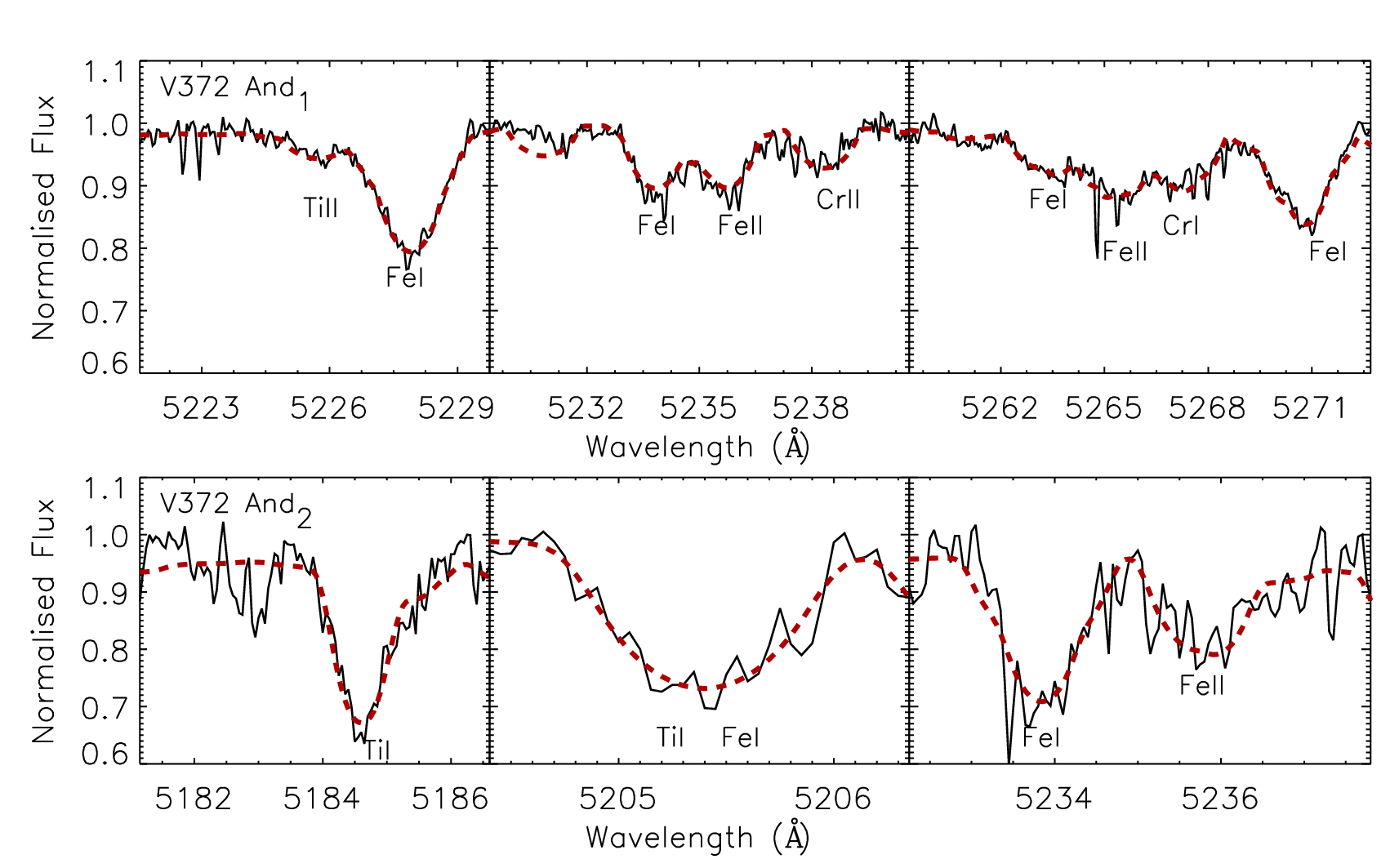}
  \end{minipage}
   \caption{Comparison of the theoretical (dashed lines) and disentangled spectra (continuous lines) of component stars of V2080\,Cyg (left panels) and V372\,And (right panels). 
   The subscripts 1 and 2 represent the primary and the secondary components, respectively.
   A similar figure for CF\,Lyn is given in Fig.\,\ref{spectrafit2}}
\end{figure*}

We first determined the initial \teff\, values of the component stars from the H$\beta$ line analysis. Before starting the analysis, we 
checked the normalisation of H$\beta$ lines by comparing them with synthetic spectra, as these broad lines are affected by the \'{e}chelle orders merging. 
If necessary a re-normalisation process was applied 
with the \textit{continuum} task of the NOAO/IRAF package\footnote{http://iraf.noao.edu/}.

In the H$\beta$ line analysis, the 
\teff\, values were obtained considering the minimization method \citep{2004A&A...425..641C}. During the analysis, the 
metallicity and \logg\, parameters were assumed to be solar and 4.0 cgs, respectively. 
% The \logg\, has no significant 
% effect on hydrogen lines for \teff\, values lower than 8000\,K (\citeauthor{smalley02}\,\citeyear{smalley02}). 
The resulting H$\beta$ \teff\, values were determined taking into account the minimum difference between the synthetic and 
observed spectra. The uncertainties in \teff\, values were estimated 
considering the 1$\sigma$ change in the difference between the observed and synthetic spectra and also taking 100\,K error introduced by 
normalisation. The results of the analysis 
are given in Table\,\ref{atmospar}. A comparison between the theoretical and disentangled H$\beta$ lines is shown in 
Fig.\,2. 

Final atmospheric parameters were determined based on the excitation and the ionization equilibrium of iron (Fe) lines. 
This is because for the accurate atmospheric parameters, the abundance of an element obtained from different excitation or ionization potential must be the same.

\begin{figure}
\centering
\includegraphics[width=8cm,height=8cm, angle=0]{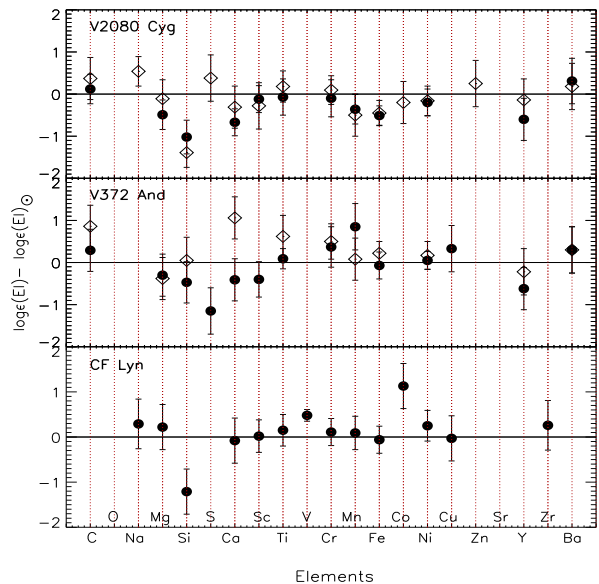}
\caption{Chemical abundances distribution of stars relative to the solar values \citep{2009ARA&A..47..481A}. The filled circle and diamond 
symbols represent the primary and the secondary components, respectively.}
\label{abundist}
\end{figure}

The disentangled spectra of the component stars were divided into small spectral parts considering the normalisation level. 
The line identification of each spectral part was made by using the Kurucz line 
list\footnote{kurucz.harvard.edu/linelists.html}. In the analysis, we used the 
spectrum synthesis method which generates the theoretical spectrum until it fits well the observed one. 
In the analysis, each spectral part was analysed 
separately.

\teff\, and $\xi$ parameters sensitively depend on the strength of \ion{Fe}{i} lines, while they show no significant dependence on \ion{Fe}{ii} lines. On the contrary, 
\logg\, parameter is very sensitive to the strength of \ion{Fe}{ii} lines. Therefore, the \teff\, values of the component stars were determined by 
considering the correlation between \ion{Fe}{i} abundance and the excitation potential. Additionally, the $\xi$ parameters were derived by checking 
the relation between the depth of \ion{Fe}{i} lines and abundances. The \logg\, values were also derived relying on the relationship between the ionization 
balance and abundances of \ion{Fe}{ii} (for more information see \citeauthor{2016MNRAS.458.2307K} \citeyear{2016MNRAS.458.2307K}). The \vsini\, values were also 
determined during the analysis. To estimate the errors in the \teff, \logg\, and $\xi$, we took into account $\sim$5\% 
changes in the relationships of the excitation potential$-$abundance, the ionization potential$-$abundance and the line depth$-$abundance, respectively. 
The derived final atmospheric parameters and the \vsini\, values of each component stars are given in Table\,\ref{atmospar}. 
The atmospheric parameters of the second and third components of CF\,Lyn and V2080\,Cyg could not be derived because of their low 
S/N ratio disentangled  spectra.

\begin{table*}
\centering
  \caption{The H$\beta$ \teff\, values, the final atmospheric parameters, \vsini\, and the Fe abundances of the component stars.
  The subscripts 1 and 2 represent the primary and the secondary components, respectively.}
  \label{atmospar}
  \begin{tabular*}{0.9\linewidth}{@{\extracolsep{\fill}}lcccccc}
%   \begin{tabular}{lccccc}
\toprule
                 &\multicolumn{1}{c}{\hrulefill H$\beta$ line\,\hrulefill}
                 &\multicolumn{5}{c}{\hrulefill \,Fe lines\,\hrulefill}\\
Star             & \teff\,\,(K)   &  \teff\,\,(K)     & \logg\,\,(cgs)     & $\xi$\,(\kms)   & \vsini\,\,(\kms) & $\log \epsilon$ (Fe)  \\
\hline
V2080\,Cyg$_{1}$   &6200\,$\pm$\,200   &6100\,$\pm$\,100   & 4.0\,$\pm$\,0.1    & 1.8\,$\pm$\,0.2 & 23\,$\pm$\,3      &6.99\,$\pm$\,0.25\\
V2080\,Cyg$_{2}$   &6100\,$\pm$\,300   &6300\,$\pm$\,200   & 4.2\,$\pm$\,0.1    & 2.1\,$\pm$\,0.2 & 22\,$\pm$\,3      &7.17\,$\pm$\,0.29\\
V372\,And$_{1}$    &7900\,$\pm$\,200   &8000\,$\pm$\,100   & 3.9\,$\pm$\,0.1    & 2.4\,$\pm$\,0.2 & 56\,$\pm$\,4      &7.43\,$\pm$\,0.25\\
V372\,And$_{2}$    &7800\,$\pm$\,300   &7800\,$\pm$\,200   & 4.2\,$\pm$\,0.2    & 2.7\,$\pm$\,0.3 & 33\,$\pm$\,3      &7.72\,$\pm$\,0.33\\
CF\,Lyn$_{1}$      &6200\,$\pm$\,300   &6600\,$\pm$\,200   & 4.0\,$\pm$\,0.1    & 2.3\,$\pm$\,0.3 & 82\,$\pm$\,5      &7.44\,$\pm$\,0.26\\
\bottomrule
\end{tabular*}
\end{table*}

After the accurate atmospheric parameters were derived, we performed the abundance analysis by taking these parameters fixed. The abundances were 
derived by using the spectrum synthesis method which is more suitable for the analysis of stars having slow to high 
\vsini\, values. The line-identified spectral parts were analysed separately. The final element abundances were determined by adjusting the 
abundances of individual elements and considering the minimum difference between the observed and synthetic spectra. After the analysis of each spectral parts of 
individual stars, the average values of the calculated element abundances were obtained.
The obtained theoretical spectra fits for the observed lines of the components of V2080\,Cyg and V372\,And are shown in Fig.\,3. 

There are a few quantities which affect the uncertainties of chemical element abundances. The assumption in model atmosphere calculations, 
quality of observed spectra (S/N and resolution), error in determined atmospheric parameters are significant uncertainty sources. To estimate 
the real error in the derived chemical abundances, we checked the uncertainty effects of those sources. It was calculated that the assumptions 
(local thermodynamical equilibrium, plane-parallel geometry, and hydrostatic equilibrium) in model atmosphere 
contribute about 0.1\,dex error in abundances \citep{2011mast.conf..314M}. The quality effect of a spectrum was examined by \citet{2016MNRAS.458.2307K}. 
We took the uncertainty contribution of this effect from this study. Additionally, we calculated by how much chemical abundance is changed with 100\,K difference in \teff, 0.1\,cgs 
variation in \logg, and 0.1\kms\, in $\xi$. It turned out that the variations in \teff\, and $\xi$ make around 0.08\,dex difference in abundance. The \logg\, error contribution 
was also found $\sim$0.05\,dex. Those results are similar to the errors found by \citet{2016MNRAS.458.2307K}. The resulting uncertainties for the elements were 
calculated considering all error contributions. The Fe line abundance and a list of the element abundances of each component star are given in Table\,\ref{atmospar} and Table\,A2, respectively. 
In Table\,A2, the number of analysed spectral parts is given in the brackets. As can be noticed, abundance values for some 
elements were obtained from less number of spectral parts. Therefore, it should be kept in mind that these abundances are not so reliable.  
The abundance distributions of component stars relative to solar abundance are shown in Fig.\,\ref{abundist}.

As a result of this analysis, we found that the components of V2080\,Cyg show slightly lower Fe abundance comparing to solar abundance
\citep{2009ARA&A..47..481A}. The primary component of V2080\,Cyg has a lower Fe abundance relative to the secondary component
but those Fe abundances agree with each other within the uncertainties. Additionally, it turned out that the component stars of V372\,And and CF\,Lyn have element abundances similar to solar
within the errors.

\begin{figure*}
\centering
\includegraphics[width=17cm, angle=0]{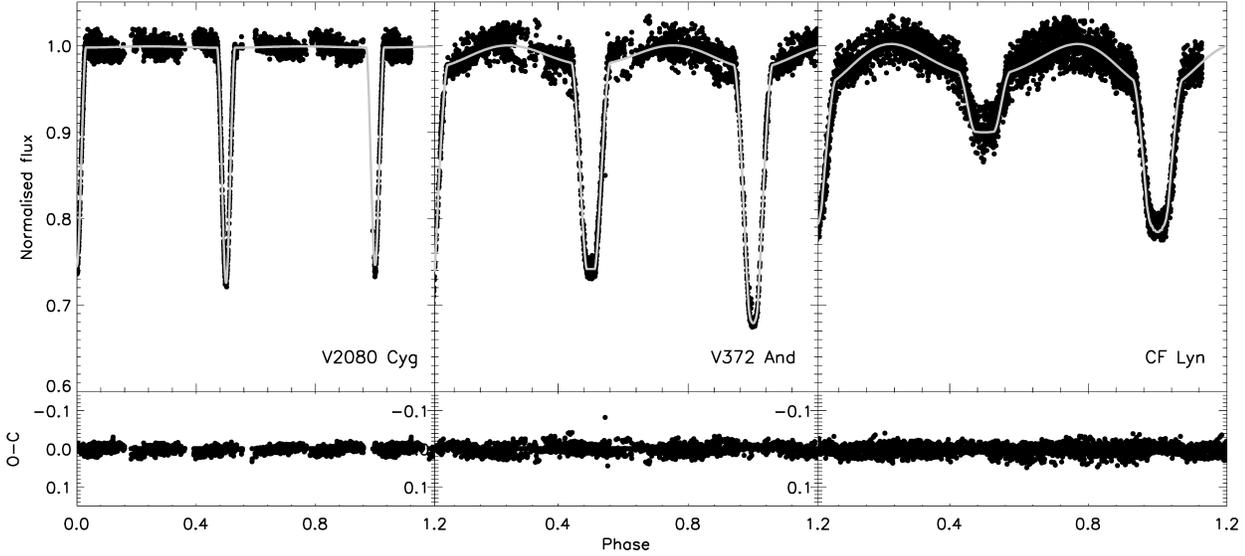}
\caption{Upper panel: comparison of observed (dots) and theoretical light curves (solid line). Lower panel: residuals.}
\label{lcfit}
\end{figure*}

\section{Light-curve solution}
The normalised SuperWASP data of the eclipsing binary systems was used in the light curve analysis.
To obtain more accurate results the SuperWASP data and the measured $v$$_{r}$ values were analysed simultaneously
by using the Wilson$-$Devinney code \citep{1971ApJ...166..605W} combined with the Monte Carlo simulation to 
calculate the uncertainties of the searched parameters \citep{2010MNRAS.408..464Z, 2004AcA....54..299Z}.
The light curves of the systems are good enough for this analysis. Only V2080\,Cyg has a lack of descending branch in the primary minimum.
However, there are enough data points to fit the primary eclipse.

Final \teff\, values of the primary components were fixed during the analysis. The bolometric albedos \citep{1969AcA....19..245R}
and the bolometric gravity-darkening coefficients
\citep{1924MNRAS..84..665V} were set to be 1 for radiative
atmospheres. They were also set as 0.5 and 0.32 for convective atmospheres, respectively. The logarithmic limb darkening coefficients were
taken from \citet{1993AJ....106.2096V} and they were also fixed during the analysis. Additionally, a synchronous rotation was assumed.
The \teff\, values of secondary components,
phase shift ($\phi$), $i$, $q$, and fractional luminosities, dimensionless potentials ($\Omega$) of component stars were set as free parameters.
The derived orbital parameters from $v$$_{r}$ analysis were used as input values. During the analysis, a detached binary configuration
was assumed. Additionally, we set the third light contribution  as a free parameter.

Consequently, we determined the final parameters of the binary systems and a third light contribution was found for 
V2080\,Cyg. Additionally, the astrophysical parameters of each 
component star were calculated by the JKTABSDIM code \citep{2004MNRAS.351.1277S} which carefully estimates the uncertainties in the derived fundamental parameters. 
The distances of the binary systems were also determined by utilizing the distance modulus. 
To calculate the distances, the bolometric correction and the absorption values were taken from \citet{2018MNRAS.479.5491E} 
and \citet{2011ApJ...737..103S}, respectively. Final orbital parameters and the derived parameters are listed in 
Table\,\ref{lcresult}. The theoretical light curves' fit to the SuperWASP light curves is demonstrated in Fig.\,\ref{lcfit}.

\begin{table*}
\begin{center}
\centering
\label{lcresult}
\caption{Results of the light curve analysis and the astrophysical parameters.
The subscripts 1, 2 and 3 represent the primary, the secondary, and third components, respectively. $^a$ shows the fixed parameters.}
\begin{tabular}{lrrr}
\hline
 Parameter			  &  V2080\,Cyg   	    &V372\,And			    & CF\,Lyn		\\	
\hline
$i$ ($^{o}$)	       	          & 86.009 $\pm$ 0.091      &86.947 $\pm$ 0.209             &83.818 $\pm$ 0.210  \\	
\teff$_{1}$$^a$ (K)               & 6100 $\pm$ 100  	    &8000 $\pm$ 100		    &6600 $\pm$ 200	\\	
\teff$_{2}$ (K)    	          & 6210 $\pm$ 250	    &7620 $\pm$ 340		    &5220 $\pm$ 260		\\
$V$$_{\gamma}$ (\kms)		  & 1.170 $\pm$ 0.321       &-2.490 $\pm$ 0.003 	    &1.364 $\pm$ 0.051  	  \\
$a$ ($R$$_{\odot}$)        	  & 16.254 $\pm$ 0.019      &13.742 $\pm$ 0.090 	    &6.912 $\pm$ 0.132 	  \\
$e$	         	          & 0$^a$        	    &0$^a$       		    &0.029$\pm$ 0.003		  \\
$\Omega$$_{1}$		          & 10.339 $\pm$ 0.179      &5.133 $\pm$ 0.017  	    &3.732 $\pm$ 0.010  	  \\
$\Omega$$_{2}$		          & 11.925 $\pm$ 0.242      &6.755 $\pm$ 0.038  	    &5.686 $\pm$ 0.024  	  \\
Phase shift             	  & 0.00001 $\pm$ 0.00001   &0.00011 $\pm$ 0.00009	    &0.00100 $\pm$ 0.00010	  \\
$q$                     	  & 0.982 $\pm$ 0.002       &0.775 $\pm$ 0.004  	    &0.586 $\pm$ 0.003  	  \\
$r$$_{1}$$^*$ (mean)              & 0.1067 $\pm$ 0.0019     &0.2306 $\pm$ 0.0010            & 0.3250 $\pm$ 0.0023  \\
$r$$_{2}$$^*$ (mean)              & 0.0898 $\pm$ 0.0018     &0.1373 $\pm$ 0.0009            & 0.1316 $\pm$ 0.00107 \\
$L$$_{1}$ / ($L$$_{1}$+$L$$_{2}$) & 0.568 $\pm$ 0.024       &0.767 $\pm$ 0.010  	    &0.943 $\pm$ 0.009  	  \\
$L$$_{2}$ /($L$$_{1}$+$L$$_{2}$)  & 0.432 $\pm$ 0.020       &0.233 $\pm$ 0.009  	    &0.057 $\pm$ 0.009 	  \\
$l$$_{3}$                         & 0.083 $\pm$ 0.015       &				    &				  \\
\multicolumn{4}{c}{Derived Quantities}\\

$M$$_{1}$ ($M_\odot$)	          & 1.197 $\pm$ 0.005  	    &2.282 $\pm$ 0.048		     &1.457 $\pm$ 0.099	\\	
$M$$_{2}$ ($M_\odot$)	          & 1.173 $\pm$ 0.004       &1.748 $\pm$ 0.035		     &0.857 $\pm$ 0.037		\\
$R$$_{1}$ ($R_\odot$)	          & 1.734 $\pm$ 0.031       &3.170 $\pm$ 0.025		     &2.247 $\pm$ 0.046		  \\
$R$$_{2}$ ($R_\odot$)		  & 1.459 $\pm$ 0.029       &1.887 $\pm$ 0.017		     &0.909 $\pm$ 0.021		  \\
log\,$L$$_{1}$ ($L_\odot$)        & 0.575 $\pm$ 0.032       &1.570 $\pm$ 0.023		     &0.937 $\pm$ 0.056		  \\
log\,$L$$_{2}$ ($L_\odot$)	  & 0.457 $\pm$ 0.072       &1.034 $\pm$ 0.078		     &-0.257 $\pm$ 0.088		  \\
\logg\,$_{1}$ (cgs)               & 4.038 $\pm$ 0.016 	    &3.794 $\pm$ 0.005  	     &3.898 $\pm$ 0.014  			    \\
\logg\,$_{2}$ (cgs)               & 4.179 $\pm$ 0.017 	    &4.129 $\pm$ 0.007		     &4.453 $\pm$ 0.009 			    \\
$M_{bolometric}$$_{1}$ (mag)      & 3.313 $\pm$ 0.081       &0.825 $\pm$ 0.057               &2.408 $\pm$ 0.139		  \\
$M_{bolometric}$$_{2}$ (mag)	  & 3.608 $\pm$ 0.180       &2.165 $\pm$ 0.195	             &5.394 $\pm$ 0.221		  \\
% $M_{V}$$_{1}$ (mag)	          & 3.341 $\pm$ 0.091       &0.805 $\pm$ 0.044	             &2.44 $\pm$ 0.14		  \\
% $M_{V}$$_{2}$ (mag)	          & 3.631 $\pm$ 0.195       &1.894 $\pm$ 0.194	             &5.61 $\pm$ 0.30		  \\
Distance (pc)                     & 86  $\pm$ 4             &472 $\pm$ 29                    &231 $\pm$ 17  			    \\
 \hline
\end{tabular}
     \end{center}
     \begin{description}
     \centering
 \item[ ] *fractional radius, $R$/$a$.
 \end{description}
\end{table*}

\section{Discussion}

\subsection{Fundamental properties}

As a result of the analysis of the light and radial velocity curves, the fundamental stellar parameters, $L$ and 
$M_{bolometric}$ values of all component stars were obtained.
The accuracy of the $M$ and $R$ values ranges between 1\% and 2\% for V2080\,Cyg and V372\,And. However,
the accuracy of these parameters reaches $4-6$\% for $M$ and $\sim$2\% for $R$ in CF\,Lyn case. As CF\,Lyn
has an eccentric orbit, the error in the determined parameters come higher comparing the other systems. To
decrease these error bars, new spectroscopic data taken from whole orbital phases is necessary. When the accuracy criteria given
by \citet{2018MNRAS.479.5491E} is considered, the derived fundamental parameters of V2080\,Cyg and V372\,And are very accurate ($M$ and $R$ uncertainties $\leq$ 3\%) and
these obtained parameters are accurate ($M$ and $R$ uncertainties 3$-$6\%) for CF\,Lyn.
The distances of systems were calculated to be 472\,$\pm$\,29, 231\,$\pm$\,17, and 86\,$\pm$\,4\,pc for V372\,And, CF\,Lyn,\, and V2080\,Cyg, respectively.
These distance values were found to be consistent with the distance values given by Gaia  \citep{2018A&A...616A...1G}.

A synchronous rotation was assumed in the light curve analysis. Therefore, the rotation period of the component stars should be the same
with the orbital period of binary system. Using this approach and the derived radii of the component stars, we can calculate the synchronous
\vsini\, parameters of components. The synchronous \vsini\, values of the primary and the secondary components of CF\,Lyn were found to
be 81 and 37 \kms, respectively. As we only obtained the good quality disentangled spectrum of the primary component of CF\,Lyn,
only the \vsini\,
value of the primary component star was derived from the spectroscopic analysis. The synchronous and the spectroscopic
\vsini\, values of CF\,Lyn are consistent with each other. For the primary and secondary components of V372\,And, the synchronous \vsini\,
values were calculated as 55 and 36 \kms. These values also agree with the spectroscopic \vsini\, within the errors. The synchronous \vsini\, values
of the primary and secondary components of V2080\,Cyg were also obtained as 18 and 15 \kms\,, respectively. These values are slightly different
than the spectroscopic \vsini\, values. This difference can be caused by the effect of the third component.

\begin{figure}
 \centering
 \label{roche}
 \begin{minipage}[b]{0.25\textwidth}
  \includegraphics[height=2cm, width=0.95\textwidth]{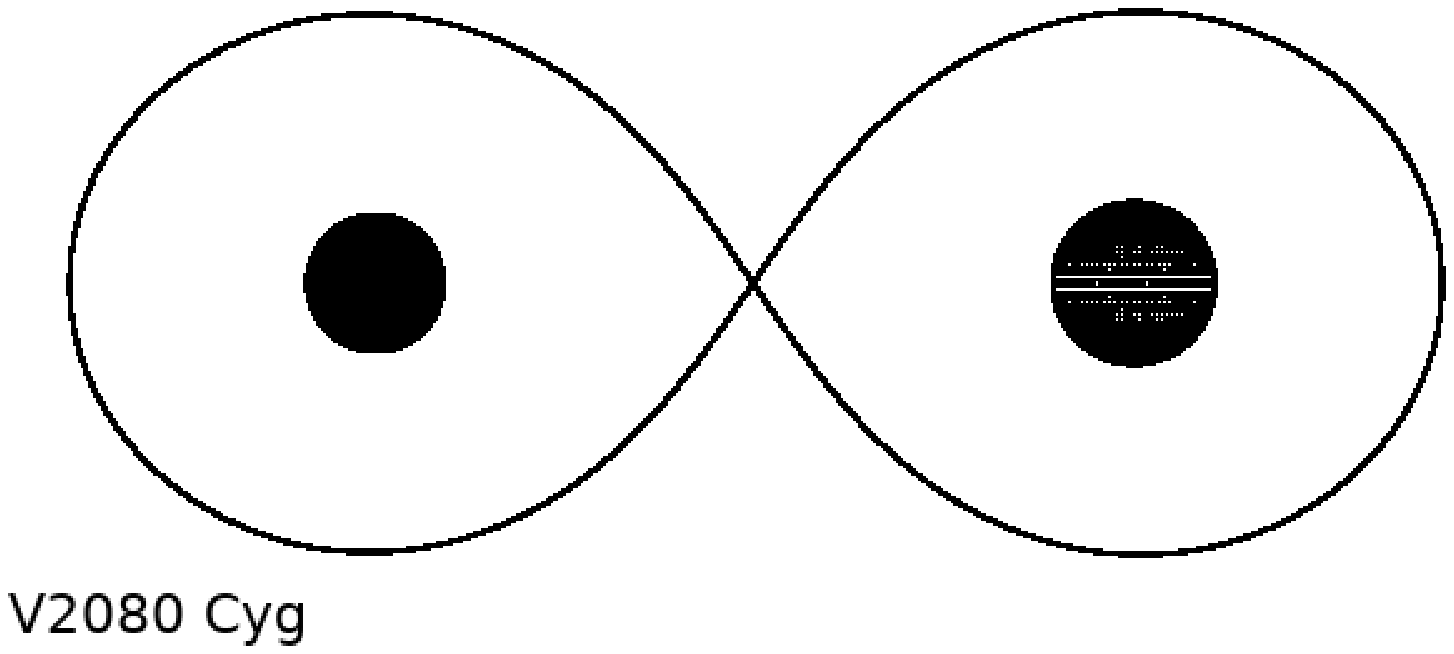}
%   \caption{1}
 \end{minipage}
 \begin{minipage}[b]{0.25\textwidth}
  \includegraphics[height=2cm, width=1\textwidth]{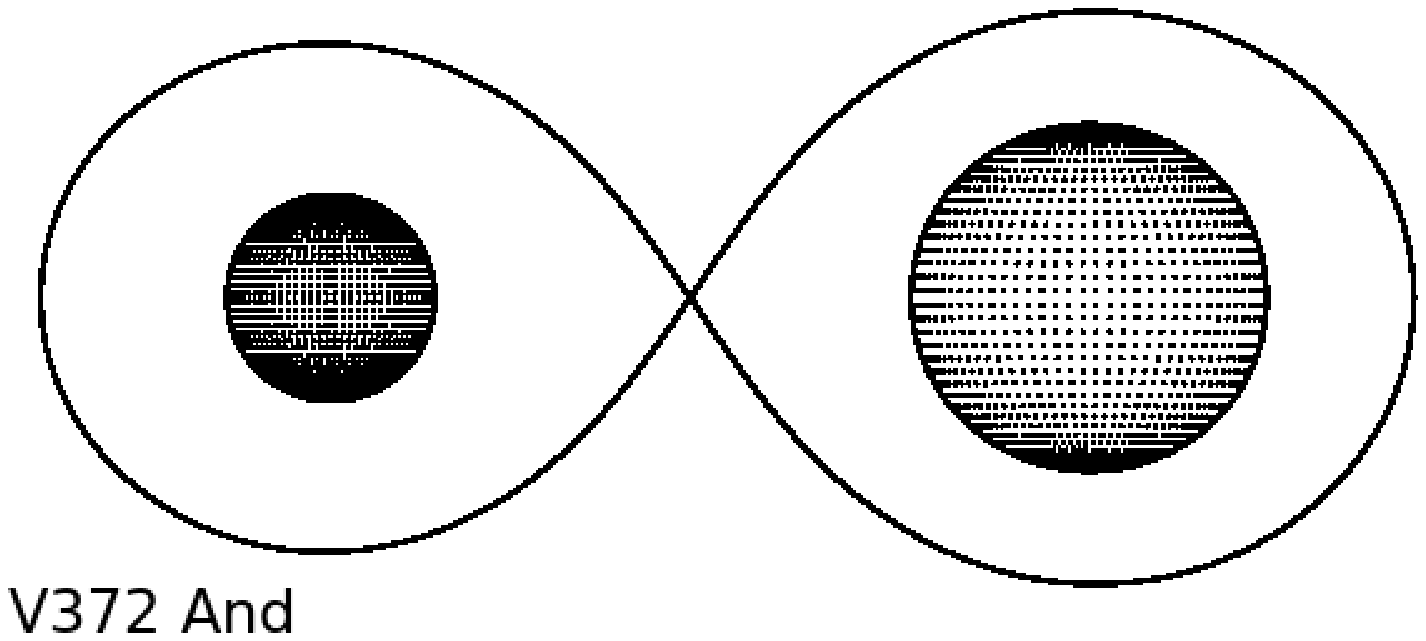}
  \end{minipage}
 \begin{minipage}[b]{0.25\textwidth}
  \includegraphics[height=2cm, width=1\textwidth]{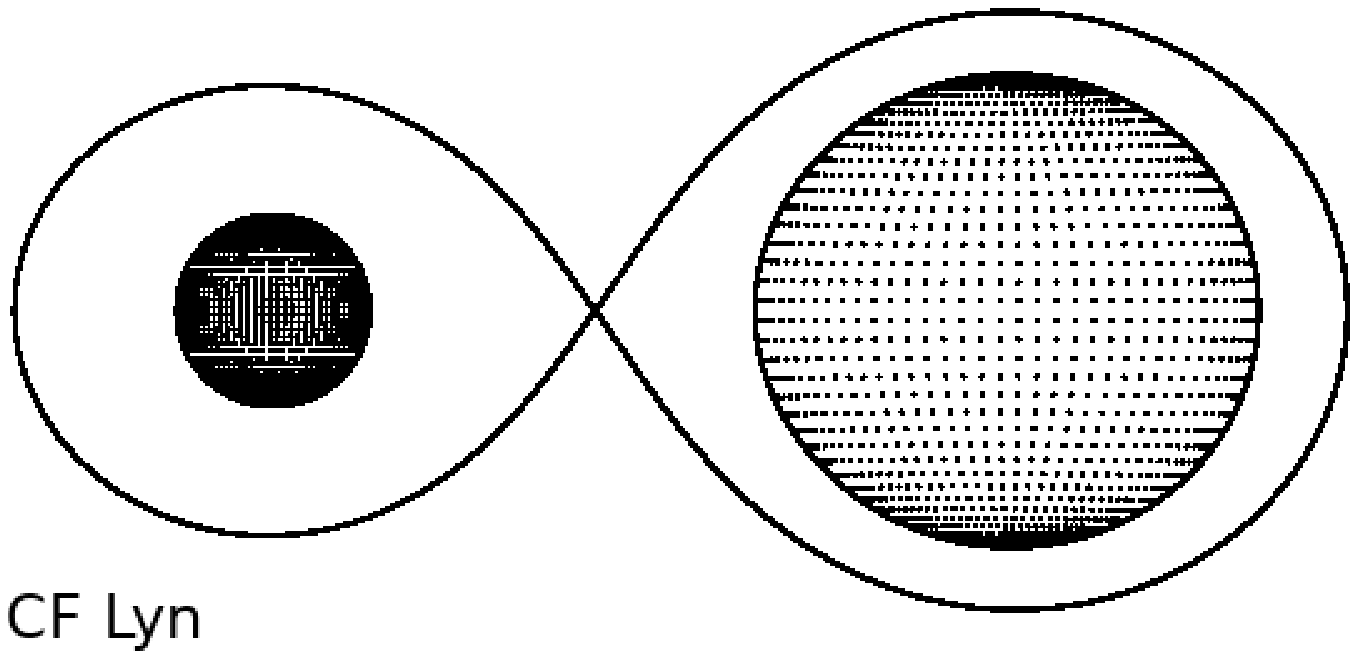}
%   \caption{3}
 \end{minipage}
   \caption{Roche geometries of the detached systems at orbital phase 0.75.}
\end{figure}

In previous studies, it was indicated that an evolution from detached to semi-detached binaries and also from semi-detached to contact
binaries exist \citep[][e.g.]{2012AcA....62..153S, 2005ApJ...629.1055Y}.
Particularly, the detached systems having the $q$ value lower than average (the average $q$ value is $\sim$0.9,
\citeauthor{2014PASA...31...24E} \citeyear{2014PASA...31...24E}) are good candidates to examine the evolution from
detached to semi-detached binaries. In our sample, the detached binary systems have different $q$ values. The $q$ values are about 0.78, 0.98 and 0.59 for V372\,And, 
V2080\,Cyg and CF\,Lyn, respectively. We calculated the Roche lobe filling factor\footnote{Filling factor\,=\,($\Omega$$_{inner}$ - $\Omega$) / ($\Omega$$_{outer}$ - $\Omega$$_{inner}$)} of each component star to see how much they fill their Roche lobes. It turned out that the primary
and secondary components of V2080\,Cyg have a filling factor of 37\% and 32\%, respectively. The filling factor value
for the primary component of V372\,And was found as 67\% and for the secondary component, it was obtained as 51\%.
Furthermore, we found that the primary and secondary components of CF\,Lyn fill
their Roche lobes in a ratio of 80\% and 54\%, respectively. The Roche geometries of the detached binary systems are illustrated in Fig.\,6.
For detached binary systems, it was shown that the component stars having the filling factor value $>$\,75\% exhibit non-spherical shapes \citep{2014PASA...31...24E}.
As demonstrated in Fig.\,6, CF\,Lyn also shows a non-spherical shape.

\subsection{Atmospheric parameters and chemical composition}

Using the disentangled spectra of component stars, we derived their fundamental atmospheric parameters, \vsini\, values and the chemical 
compositions by applying the spectrum synthesis method. The \teff\, values of components were determined from H$\beta$ and Fe lines. The 
obtained \teff\, value of V2080\,Cyg is consistent with the assumed \teff\, used in previous light curve analysis 
\citep{2008MNRAS.384..331I}. The \logg\, values were also calculated using the derived parameters from the light curve analysis 
and those values were found to be consistent with the spectroscopic \logg\, values within the errors. 

The element abundances were determined as well. The Fe abundances of each component stars are given in Table\,\ref{atmospar}. 
As can be seen from the table, the components of CF\,Lyn and V372\,And have Fe abundance similar to solar (7.50\,dex). We found that the 
components of V2080\,Cyg show slightly lower Fe abundance comparing to solar and they have similar Fe abundance within the error bars which is 
expected in the theory since 
the binary stars are assumed to form in the same interstellar area.  
We also found that the secondary component of V2080\,Cyg is slightly hotter than the primary one. 

% According to the theory of stellar evolution, massive stars evolve faster. 
% However, although both components of V2080\,Cyg have similar $M$ and \teff\, values within error bars, the secondary 
% component star has bigger radius and it fills its Roche lobe more than the primary massive star. The secondary component of V2080\,Cyg seems to evolve faster than the primary 
% star. This situation of the secondary component can be explained by it's 
% low metallicity structure found in this study. 

\subsection{Evolutionary models}

The evolutionary models of the analysed detached binary systems were computed by using the Modules for Experiments in Stellar Astrophysics (MESA)
evolution code \citep{2011ApJS..192....3P, 2013ApJS..208....4P}.
MESA includes a binary module \citep{2015ApJS..220...15P} for simulating the binary orbital evolution and
finding the initial parameters of binary systems.

We calculated many evolutionary models for different starting parameters using the MESA.
The metallicity ($Z$) parameters was taken as 0.02 for examining the evolutionary status of the components 
of CF\,Lyn and V372\,And. Because when the error bars in the \ion{Fe}{} abundances of individual components are taken into account (see, Table\,3), except for the 
components of V2080\,Cyg, the other components have \ion{Fe}{} abundance similar to solar. The components of V2080\,Cyg have 
slightly lower \ion{Fe}{} abundance. As binary stars are assumed to form in the same environment, we determined the same 
$Z$ value of 0.017 for the components of V2080\,Cyg from the best fit evolutionary model.
% These $Z$ values were estimated from the determined average Fe abundance in this study. 
Evolutionary models were analysed by comparing the $R$ and \logg\, values of the component stars.
During the analysis, orbital period and eccentricity ($e$) planes were investigated for CF\,Lyn as
it has an eccentric orbit. Additionally, the obtained fundamental stellar parameters of each component star were used
as initial parameters in the analysis. As a result,
the most probable ages were determined and the orbital evolution scenarios were produced for each system.
The best-fit evolutionary tracks and the position of each component star in the examined binary systems
are shown in Fig.\,7. The positions of the stars in the Age$-$log\,$R$ diagram are also illustrated in Fig.\,8.
The result of evolutionary models of the binary systems is given
in Table\,\ref{evoltable}.

\begin{table}
\centering
\caption{Results obtained from the best-fit evolutionary models. Assumed parameters are illustrated by * symbol.}
  \label{table1}
\begin{tabular*}{0.95\linewidth}{@{\extracolsep{\fill}}lccc}
\hline
  Parameter                & V2080\,Cyg       & V372\,And      &  CF\,Lyn   \\
                           &                  &                &            \\
\hline
$P$$_{initial}$  (days)    & 5.17 (25)     & 3.20 (8)        &  3.40 (10)                  \\
$v$$_{1 initial}$ (\kms)& 24 (5)        & 175 (25)        &  112 (15)               \\
$v$$_{2 initial}$ (\kms)& 22 (5)        & 37 (8)          &  10 (7)                 \\
$e$$_{initial}$            & 0*             & 0*               &  0.53 (5)                 \\
Age (Gyr)                  &3.95 (20) & 0.58 (4) &  2.00 (20)                 \\
\hline
% \bottomrule
\end{tabular*}
\label{evoltable}
\end{table}

For V2080\,Cyg, we obtained a younger age comparing the age value given by \citet{2008MNRAS.384..331I}.
Additionally, in all calculated evolutionary models for V2080\,Cyg, we found that the system should show a synchronous rotation.
However, the obtained spectroscopic \vsini\, values indicate that the observed situation is different.
This difference can be caused by the effect of third body in the binary system of V2080\,Cyg.
When the obtained results for V372 And are examined, it is clear that the primary component needs to start its evolution with a
very high initial \vsini\, values. For CF Lyn system, it was found that the $e$ should start at a very high value like 0.53 in the beginning of the evolution in order to be able to
adhere to the present age of the system. However, precise light curve observations are needed to examine this situation more clearly.

The components of V2080\,Cyg and V372\,And also show consistency with the models illustrated in the Age$-$log\,$R$ diagram. However, the secondary
component of CF\,Lyn differs from the model, while primary star is consistent with the model within the errors.
The obtained radius of the small mass component of CF Lyn is considerably larger than the predicted values ​​from the models.
The most likely explanation for this situation may be the result of the magnetic activity of the small mass component
\citep{2008AJ....136.2158T, 2009arXiv0902.2548S}.
In addition to the the magnetic activity, the other factor can be the possibility of periodic mass transfer between the components as a result of the
evolution of the orbit which started with a high degree of eccentricity. This may cause the discrepancy in the surface abundance and the radius of the low mass component.

\begin{figure}
 \label{ageR}
 \begin{minipage}[b]{0.25\textwidth}
  \includegraphics[height=5.3cm, width=1.7\textwidth]{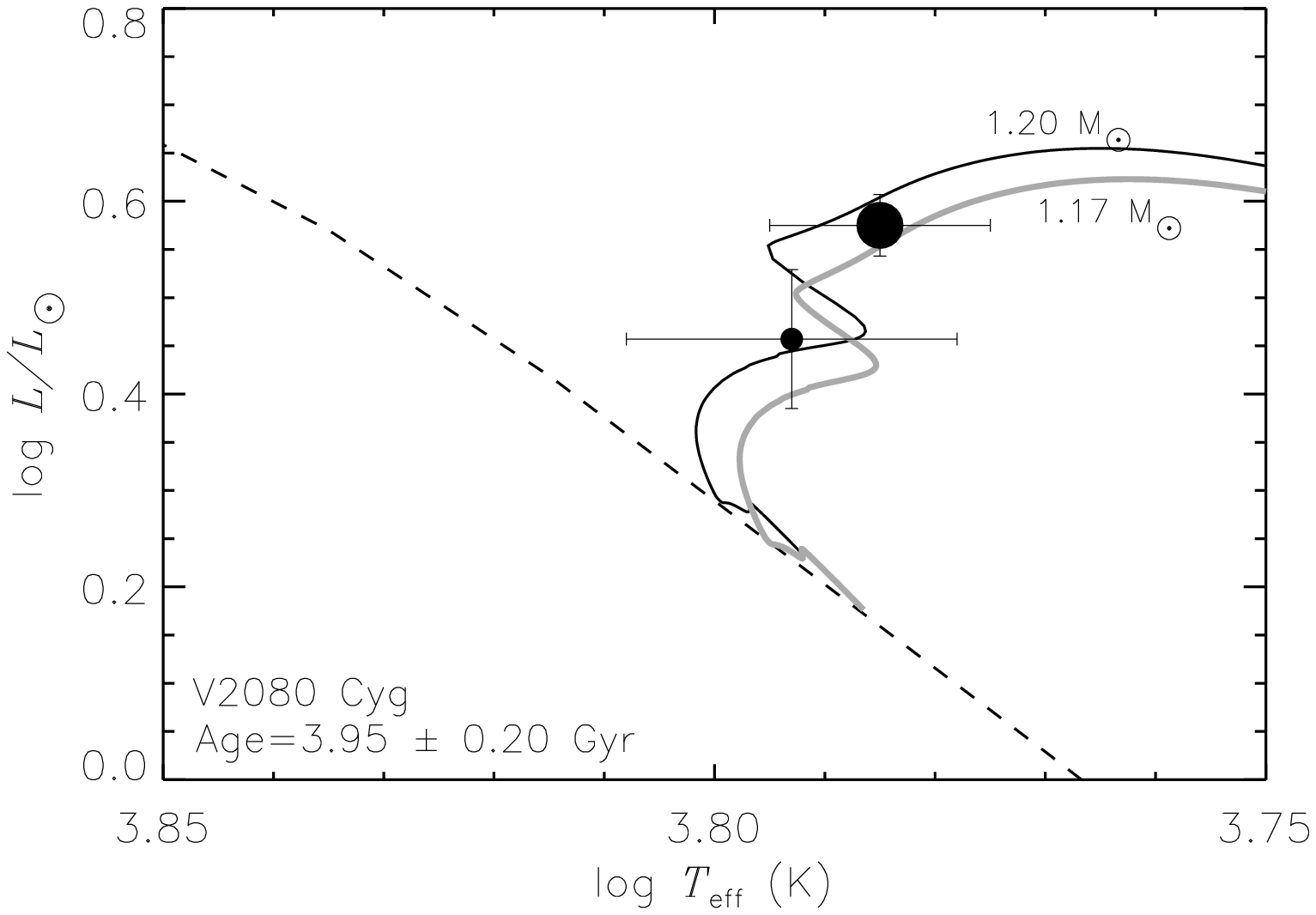}
%   \caption{1}
 \end{minipage}
 \begin{minipage}[b]{0.25\textwidth}
  \includegraphics[height=5.3cm, width=1.7\textwidth]{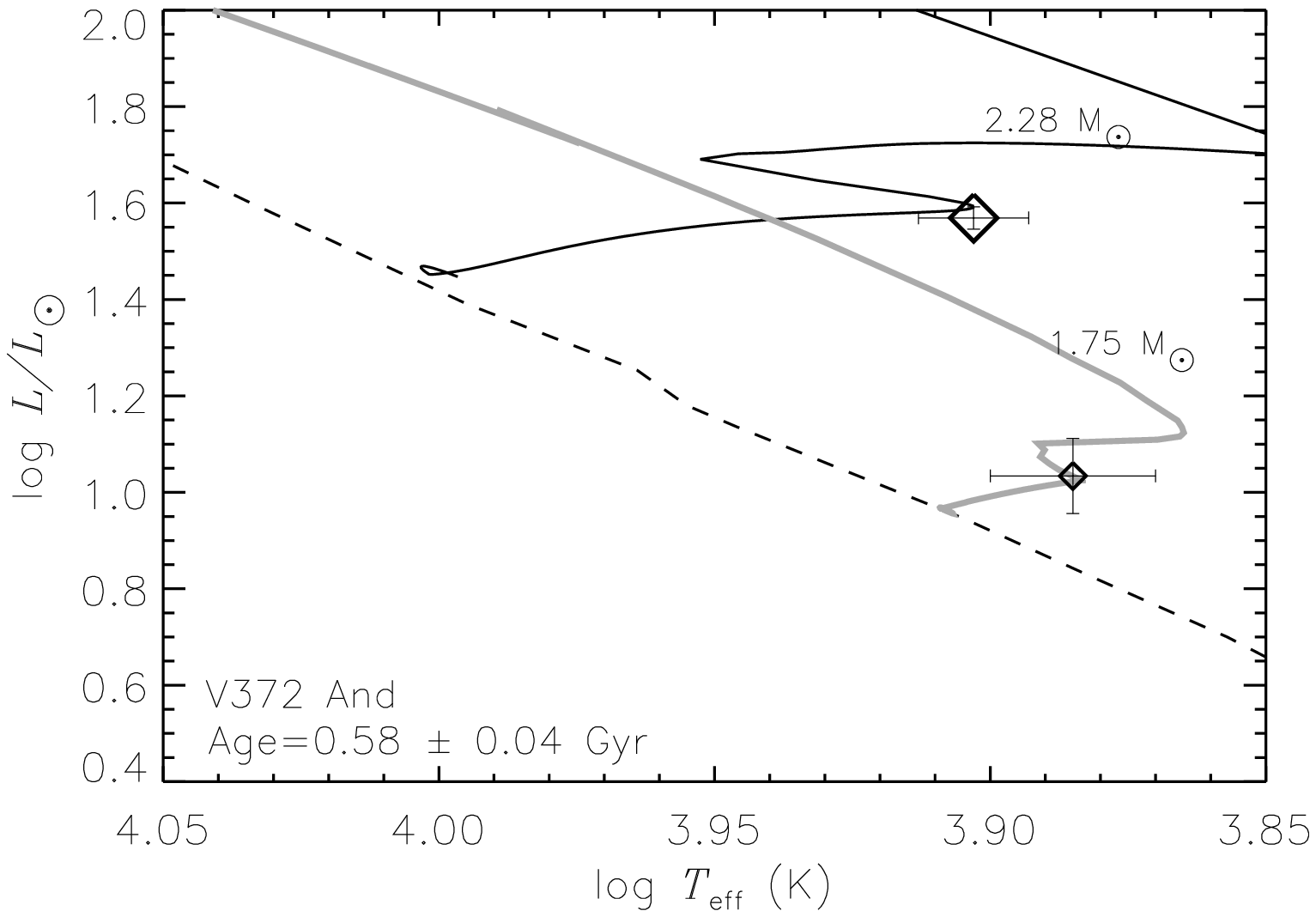}
  \end{minipage}
 \begin{minipage}[b]{0.25\textwidth}
  \includegraphics[height=5.3cm, width=1.7\textwidth]{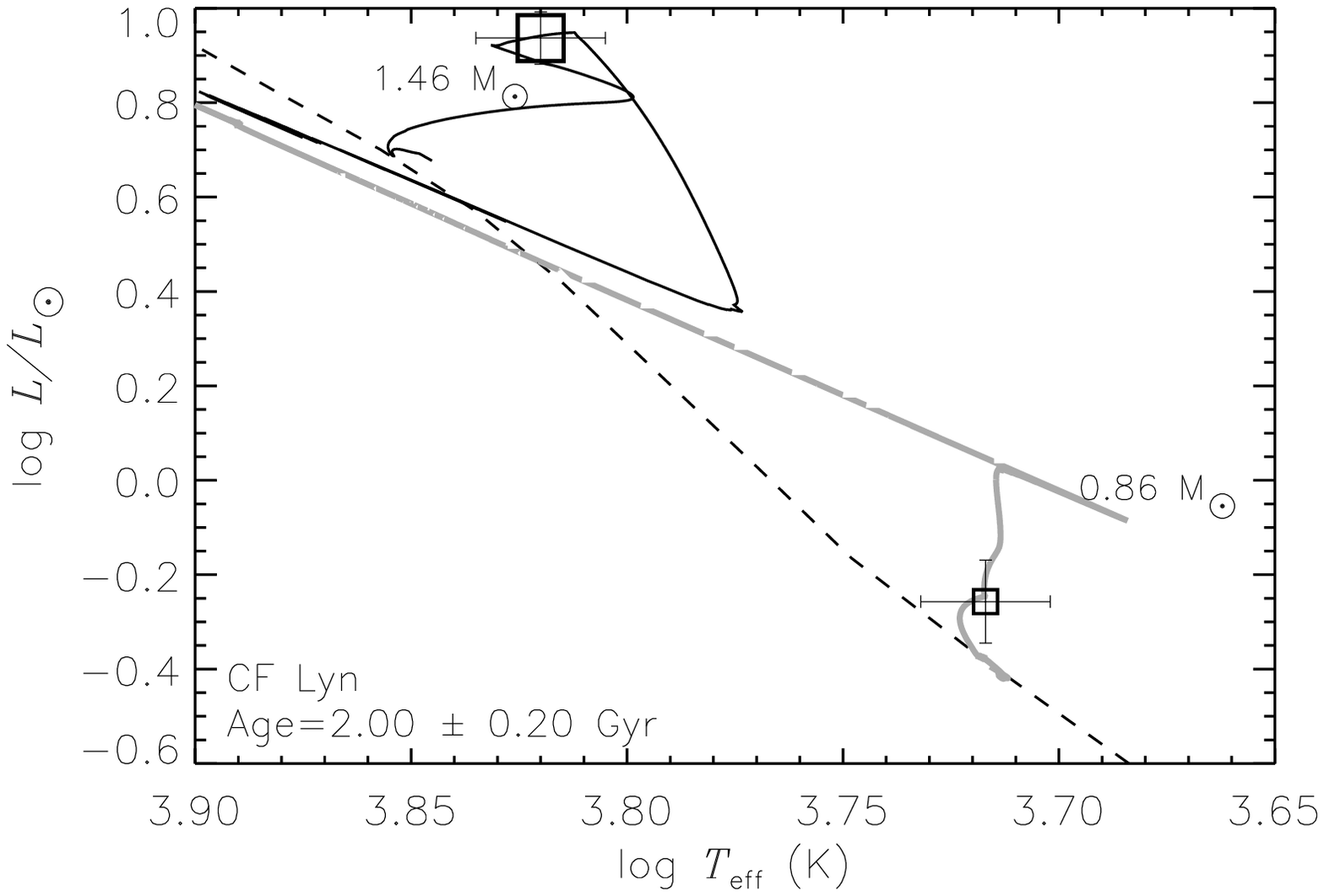}
%   \caption{3}
 \end{minipage}
   \caption{Position of the primary and the secondary (smaller symbols) component stars on the Hertzsprung-Russell diagram.
Dashed line represents the zero age main sequence (ZAMS). The evolutionary tracks are showed by the solid black and gray 
lines for the primary and secondary stars, respectively. The $Z$ value is 0.02 for CF\,Lyn and V372\,And, while it is 0.017 for V2080\,Cyg.}
\end{figure}

% % % % % % % % % % % % % % % % % % 

\begin{figure}
 \label{ageR}
 \begin{minipage}[b]{0.25\textwidth}
  \includegraphics[height=5.2cm, width=1.7\textwidth]{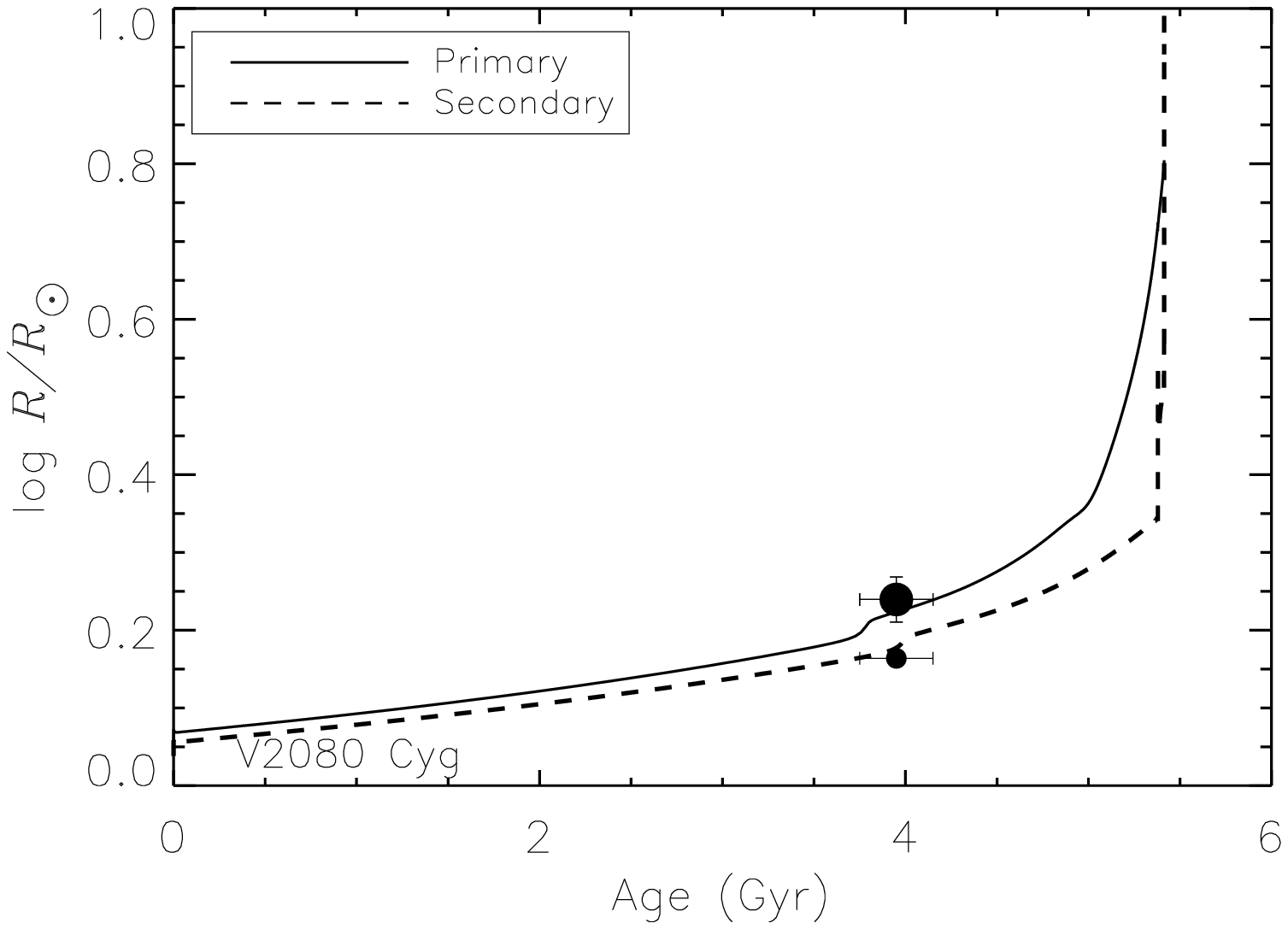}
%   \caption{1}
 \end{minipage}
 \begin{minipage}[b]{0.25\textwidth}
  \includegraphics[height=5.3cm, width=1.7\textwidth]{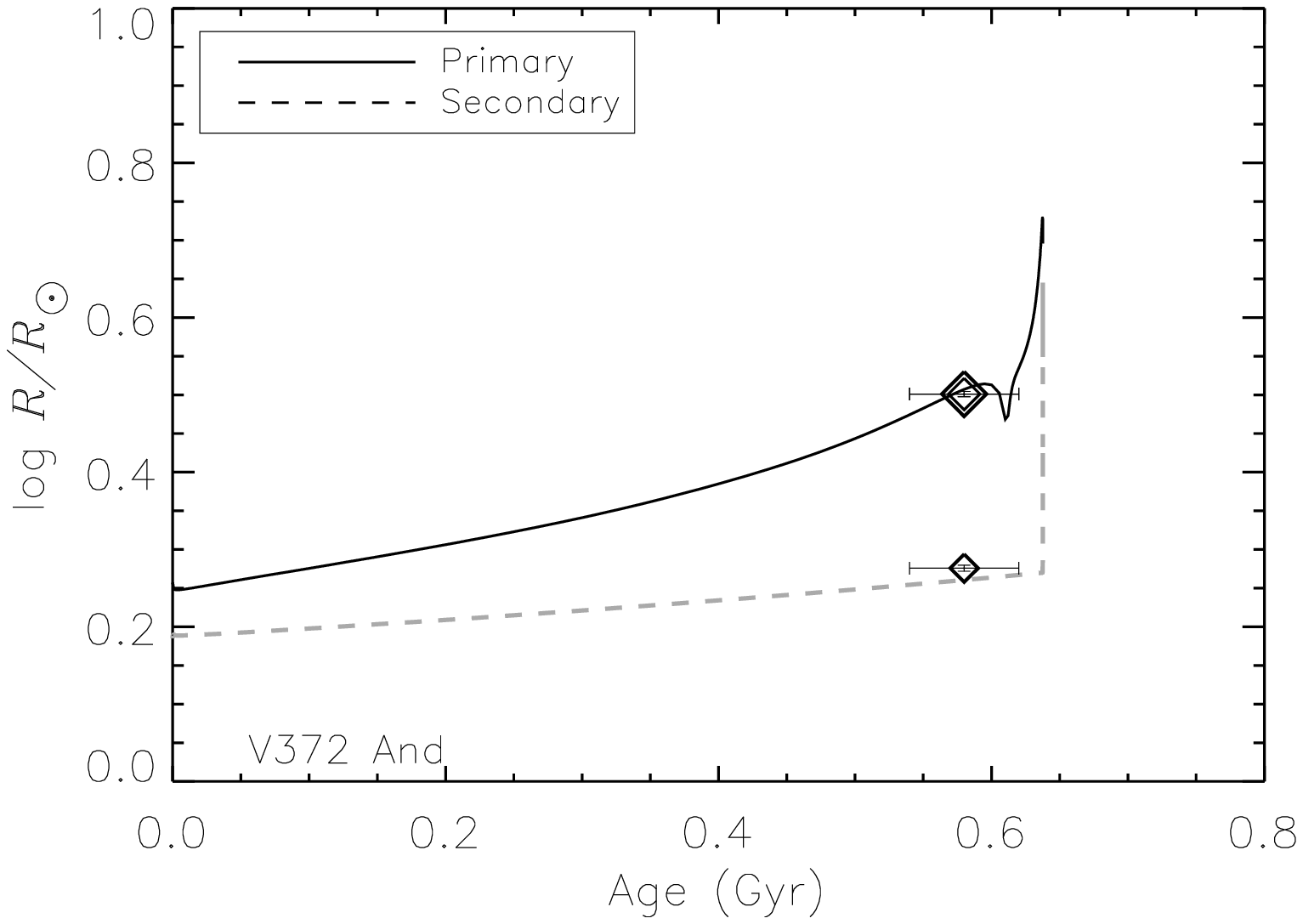}
  \end{minipage}
 \begin{minipage}[b]{0.25\textwidth}
  \includegraphics[height=5.3cm, width=1.7\textwidth]{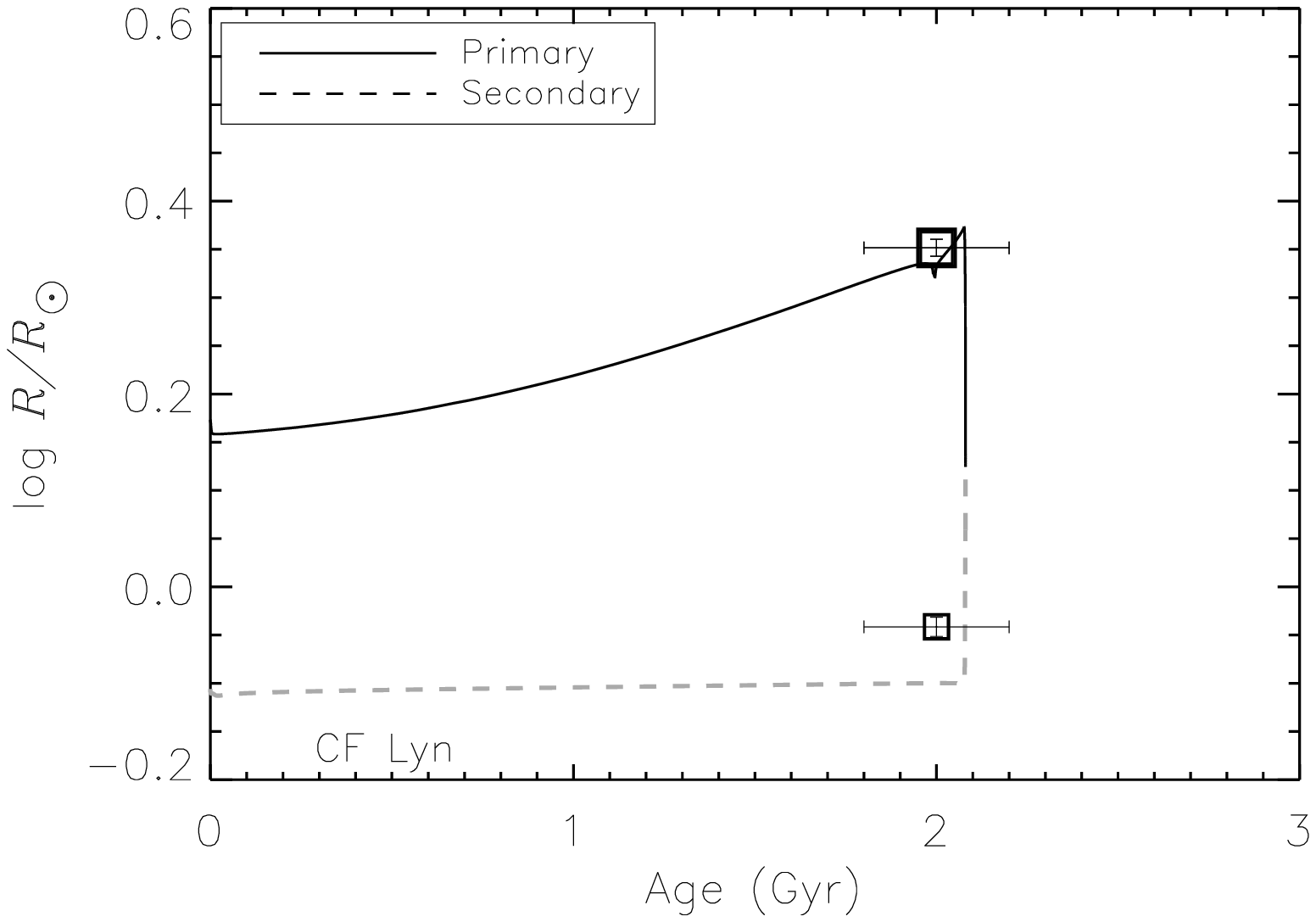}
%   \caption{3}
 \end{minipage}
   \caption{Position of the primary and the secondary (smaller symbol) component stars in the Age$-$log\,$R$ diagram. 
   The tracks are showed by the solid black and gray 
lines for the primary and secondary (smaller symbols) stars, respectively. The $Z$ value is 0.02 for CF\,Lyn and V372\,And, while it is 0.017 for V2080\,Cyg.}
\end{figure}

\section{conclusions}

In this study, a detailed examination of three double-lined detached eclipsing binary systems is presented.
We first carried out a radial velocity and light curve analysis and derived the accurate fundamental stellar parameters
of the systems. The Roche geometries and the filling factors were examined using the derived stellar parameters.
It turned out that the primary component of CF\,Lyn fills 80\% of its Roche lobe and it has a non-spherical
shape. CF\,Lyn can be a good candidate system to investigate the evolution between detached to semi-detached
binaries.

The atmospheric parameters, \vsini, and the abundances of each component star were determined by using the disentangled spectra of the 
individual systems. As a result, we found a similar \teff\, values from the H$\beta$ and Fe lines analysis within the error bars.
The spectroscopic \logg\, values were also found to be consistent with the \logg\, values calculated from the stellar parameters.
All component stars have a moderate \vsini\, value which ranges from 21 to 82 \kms. Additionally, the spectroscopic \vsini\,
values were found to agree with the synchronous \vsini\, value for CF\,Lyn and V372\,And. However, in the case of V2080\,Cyg, they differ from each other. This difference can be explained by the effect of the third component present in the system.

The chemical compositions of the components were derived. The Fe abundance of CF\,Lyn and V372\,And component stars were
obtained similar to solar within the errors. However, we found that the components of V2080\,Cyg have a lower Fe abundance
($\lesssim$0.45\, dex) comparing to the solar value. 
% This low metal abundance explains the secondary component's more evolved structure
% comparing the primary one,
% although it has almost the same mass and \teff\, values with the primary component.

\begin{figure}
\centering
\includegraphics[width=8cm, angle=0]{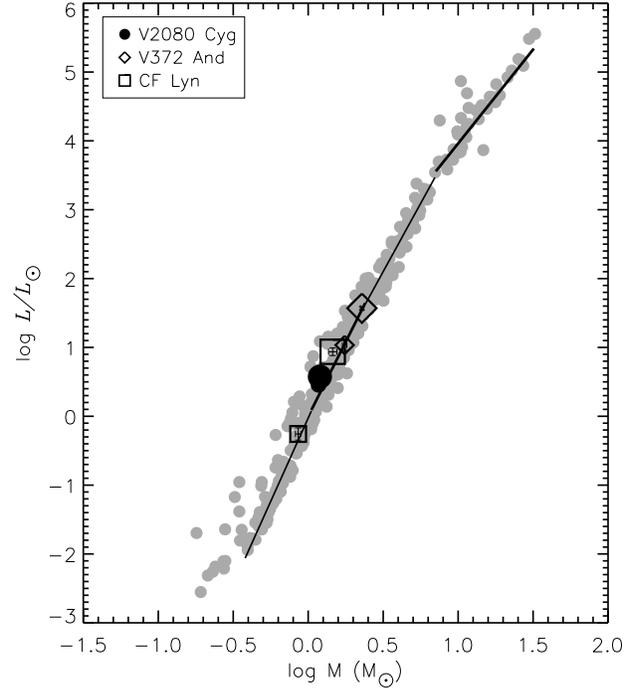}
\caption{The position of the primary and the secondary (smaller symbols) component stars in the the $\log$\,$L$\,$-$\,$\log$\,$M$ diagram.
Gray circles represent the double-lined detached eclipsing binaries given by \citet{2014PASA...31...24E}. 
The solid lines illustrate the relationships between the $L$ and $M$ \citep{2014PASA...31...24E}.}
\end{figure}

The evolutionary status of component stars were examined with the binary evolutionary models and the ages of the systems were determined.
The components of V2080\,Cyg fit well with $Z$\,=\,0.017 models within the error bars. 
In the binary evolutionary models of V2080\,Cyg, we found that the synchronous rotation
should have happened before the determined age. To explain this difference between the obtained result and the model, it is necessary
to determine the properties of the third component and investigate the orbital evolution with n-body simulations.

CF Lyn is an interesting target with an eccentric orbit. A possible explanation for the current eccentricity of
CF Lyn is may be the disruption of the orbital dynamics of the binary system by the Kozai mechanism of an invisible third body.
It was also found that the dynamic and model predicted radii of the secondary component of CF\,Lyn differ from each other. This
situation can be explained by a possible magnetic activity of the star.
For investigating such situation, we need a long-term precise light curve and high-resolution spectra with high S/N ratios.

In Fig.\,9, the position of the examined detached eclipsing binaries is shown in the $\log$\,$L$\,$-$\,$\log$\,$M$ diagram with the given double-lined
detached eclipsing binary systems by \citet{2014PASA...31...24E}. The relationship between the $L$ and $M$ of detached binaries was expressed in
four different regions taking into account the mass range \citep{2014PASA...31...24E}. These relationships are also demonstrated
in Fig.\,9. As can be seen from the figure,V2080\,Cyg, CF\,Lyn and V372\,And are consistent with the given correlations. 

We also examined whether the component stars exhibit pulsation or not. By removing the binarity effect from the
light curves a frequency analysis was performed using the Period04 \citep{2005CoAst.146...53L}. No significant result was found. However, the
light curves of the systems are not good enough to find a pulsation variability. In near future, the systems will be observed by The 
Transiting Exoplanet Survey Satellite (TESS) which will allow us to check the pulsation variability of the stars and to improve 
the light curve analysis  with its high quality photometric data.

To deeply understand the binaries, a combination of the comprehensive spectroscopic and photometric studies are necessary.
In this study, we investigated three detached eclipsing binary systems with different $q$ values.
The detailed study of these systems will help us to understand different phenomena in
binaries and their evolution. Therefore, similar analyses will improve our knowledge about the binary evolution.

\section*{Acknowledgments}
The authors would like to thank the reviewer for useful comments and suggestions that helped to improve the publication. 
We thank  \c{C}anakkale Onsekiz Mart University Research
Foundation (Project No. FBA$-$2018$-$2452) for supporting
this study. FKA thanks the Polish National Center for Science (NCN) for 
supporting the study through  grant  2015/18/A/ST9/00578.
We thank Prof. F. Soydugan for his helpful suggestions.
We are grateful to Dr. E. Niemczura for putting her codes at our disposal.  
The spectroscopic calculations have been carried out in Wroc{\l}aw Centre for Networking and Supercomputing 
(http://www.wcss.pl), grant No.\,214.
This paper makes use of data from the first public release of the WASP data \citep{2010A&A...520L..10B} as provided 
by the WASP consortium and services at the NASA Exoplanet Archive, which is operated by the California Institute 
of Technology, under contract with the National Aeronautics and Space Administration under the Exoplanet Exploration Program.
This work has made use of data from 
the European Space Agency (ESA) mission Gaia (http://www.cosmos.esa.int/gaia), processed by the Gaia Data Processing 
and Analysis Consortium (DPAC, http://www.cosmos.esa.int/web/gaia/dpac/consortium). Funding for the DPAC has been 
provided by national institutions, in particular the institutions participating in the Gaia Multilateral Agreement. 
This research has made use of the SIMBAD data base, operated at CDS, 
Strasbourq, France.

\appendix

\begin{table*}
\begin{center}
\centering
\caption[]{The $v$$_{r}$ measurements. The subscripts ``1'' and ``2'' represent the primary and the secondary component, respectively.}\label{rvmeasuremnt}

%%Please Capitalize the First Letter of Each Notional Word in table's caption
\begin{tabular*}{0.95\linewidth}{@{\extracolsep{\fill}}lcc|ccc|ccc}
% \begin{tabular}{lccc}
 \hline
\hline
HJD          & V2080\,Cyg &               & HJD       &V372\,And    &             &HJD        & CF\,Lyn     & \\
(2450000+)   &            &               &(2450000+) &             &             &(2450000+) &              &  \\
\hline
             & $v$$_{r,1}$ & $v$$_{r ,2}$ &           &$v$$_{r,1}$  & $v$$_{r,2}$ &           & $v$$_{r,1}$  &$v$$_{r,2}$ \\
\hline             
1405.3532    &-78.4\,(4)   & 85.8\,(7)    &1930.2801  & 72.9\,(1.4) &-90.0\,(2.3) &1930.4271  & 90.7\,(2.3)  &-150.8\,(5.7)\\
1407.4062    & 77.7\,(5)   &-77.0\,(6)    &1931.2977  &-99.5\,(2.7) &133.6\,(3.4) &1931.3488  &-11.1\,(1.3)  &\\ 
1408.4292    & 51.0\,(4)   &-48.7\,(5)    &1931.4540  &-99.0\,(3.0) &139.3\,(2.2) &2297.5879  & 84.4\,(1.7)  &-133.5\,(16.4)\\
1743.4947    & 76.4\,(4)   &-75.5\,(6)    &2299.3294  &-76.9\,(3.3) &103.3\,(3.3) &2299.4022  &-75.1\,(2.4)  &127.2\,(11.6)\\
2042.5990    &-26.4\,(4)   & 32.2\,(5)    &2300.3614  &100.9\,(1.5) &-126.1\,(3.3)&2300.4644  & 47.6\,(1.7)  &\\
2043.5271    & 60.0\,(5)   &-56.1\,(7)    &2302.4076  &-49.6\,(2.5) & 68.7\,(2.7) &2302.4815  &-62.2\,(1.1)  &109.3\,(10.7)\\
2490.3749    &-75.7\,(4)   & 82.2\,(6)    &2303.3568  &102.3\,(2.7) &-128.3\,(3.3)&2303.4943  &-58.0\,(1.9)  &101.0\,(9.2)\\
             &             &              &2490.5689  &-66.2\,(1.6) & 87.4\,(2.1) &2303.6085  &-86.0\,(1.6)  &145.0\,(13.5)\\
             &             &              &2491.5618  &103.7\,(2.3) &-128.5\,(3.1)&           &              &\\
 \hline
\end{tabular*}
     \end{center}
\end{table*}

\begin{figure}
\includegraphics[width=8.5cm, angle=0]{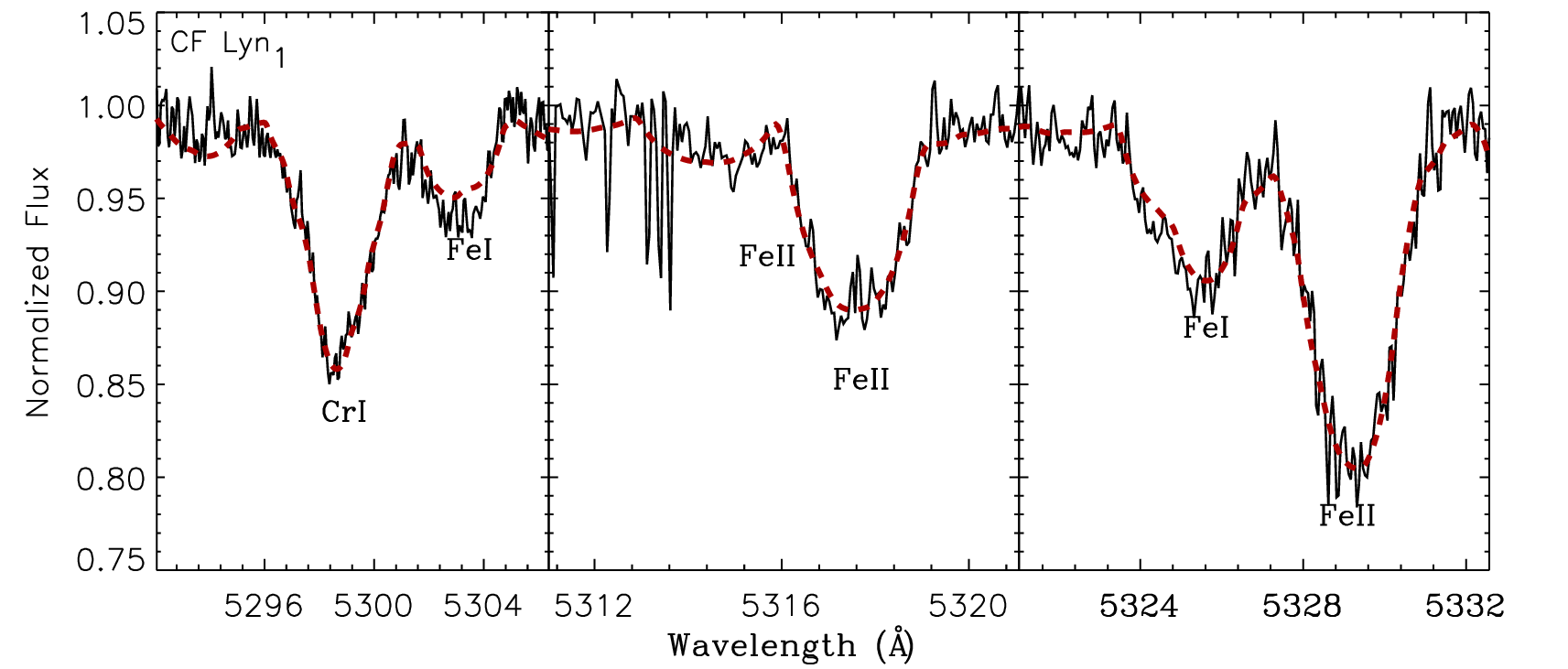}
\caption{Comparison of the theoretical (dashed line) and observed spectra (continuous line) of primary component of CF\,Lyn.}
\label{spectrafit2}
\end{figure}

\begin{table*}
\centering
\label{abundancetable}
\caption{Abundances of individual elements of the component stars. Number of the analysed spectral parts is given in the brackets.}
\begin{tabular*}{0.9\linewidth}{@{\extracolsep{\fill}}llllll}
\hline
  Elements& V2080\,Cyg$_{1}$       & V2080\,Cyg$_{2}$  & V372\,And$_{1}$        &  V372\,And$_{2}$       & CF\,Lyn$_{1}$ \\
\hline
$_{6}$C   &9.31\,$\pm$\,0.35 (1)  & 8.80\,$\pm$\,0.35 (1) &  8.72\,$\pm$\,0.37 (2) &  8.29\,$\pm$\,0.45 (2) & \\
$_{11}$Na &                       & 7.49\,$\pm$\,0.35 (1) &                        &                        & 6.53\,$\pm$\,0.40 (1)\\
$_{12}$Mg &7.76\,$\pm$\,0.35 (3)  & 7.88\,$\pm$\,0.31 (2) &  7.30\,$\pm$\,0.37 (2) &  7.22\,$\pm$\,0.45 (2) & 7.82\,$\pm$\,0.38 (2)\\
$_{14}$Si &7.12\,$\pm$\,0.32 (5)  & 7.12\,$\pm$\,0.30 (4) &  7.04\,$\pm$\,0.35 (7) &  7.56\,$\pm$\,0.46 (1) & 6.30\,$\pm$\,0.38 (2)\\
$_{20}$Ca &6.33\,$\pm$\,0.32 (5)  & 6.03\,$\pm$\,0.31 (7) &  5.93\,$\pm$\,0.37 (2) &  7.40\,$\pm$\,0.45 (2) & 6.29\,$\pm$\,0.38 (2)\\
$_{21}$Sc &3.25\,$\pm$\,0.32 (4)  & 2.87\,$\pm$\,0.32 (2) &  2.75\,$\pm$\,0.35 (4) &                        & 3.17\,$\pm$\,0.36 (3)\\
$_{22}$Ti &4.88\,$\pm$\,0.32 (17) & 5.05\,$\pm$\,0.31 (17)&  5.04\,$\pm$\,0.30 (25)&  5.57\,$\pm$\,0.40 (8) & 5.10\,$\pm$\,0.30 (24)\\
$_{23}$V  &                       & 3.84\,$\pm$\,0.30 (2) &                        &   		            & 4.41\,$\pm$\,0.35 (3)\\
$_{24}$Cr &5.68\,$\pm$\,0.26 (14) & 5.63\,$\pm$\,0.30 (23)&  6.01\,$\pm$\,0.32 (24)&  6.14\,$\pm$\,0.38 (11)& 5.75\,$\pm$\,0.30 (17)\\
$_{25}$Mn &5.20\,$\pm$\,0.32 (3)  & 4.93\,$\pm$\,0.33 (3) &  6.28\,$\pm$\,0.37 (2) &  5.51\,$\pm$\,0.45 (2) & 5.52\,$\pm$\,0.35 (9)\\
$_{26}$Fe &7.30\,$\pm$\,0.29 (71) & 7.05\,$\pm$\,0.25 (75)&  7.43\,$\pm$\,0.25 (48)&  7.72\,$\pm$\,0.33 (29)& 7.44\,$\pm$\,0.26 (56)\\
$_{27}$Co &4.83\,$\pm$\,0.35 (2)  & 4.79\,$\pm$\,0.34 (4) &  6.27\,$\pm$\,0.33 (9) &  6.39\,$\pm$\,0.42 (3) & 6.12\,$\pm$\,0.38 (2)\\
$_{28}$Ni &6.24\,$\pm$\,0.32 (15) & 6.06\,$\pm$\,0.31 (15)&  6.27\,$\pm$\,0.33 (9) &  6.39\,$\pm$\,0.42 (3) & 6.47\,$\pm$\,0.30 (22)\\
$_{29}$Cu &                       &                       &  4.52\,$\pm$\,0.38 (1) &                        & 4.16\,$\pm$\,0.38 (2)\\
$_{30}$Zn &4.45\,$\pm$\,0.35 (1)  & 4.81\,$\pm$\,0.35 (1) &                        &                        &\\
$_{38}$Sr &                       &                       &                        &                        &\\
$_{39}$Y  &2.17\,$\pm$\,0.34 (2)  & 2.07\,$\pm$\,0.35 (2) &  1.59\,$\pm$\,0.37 (2) &  1.99\,$\pm$\,0.45 (1) &\\
$_{40}$Zr &                       &                       &		           &                        & 2.84\,$\pm$\,0.40 (1)\\
$_{56}$Ba &1.69\,$\pm$\,0.35 (1)  & 2.36\,$\pm$\,0.35 (2) &  2.48\,$\pm$\,0.38 (1) &  2.48\,$\pm$\,0.45 (1) &\\
\hline
% \bottomrule
\end{tabular*}
\label{abunresult}
\end{table*}

\end{document}